\documentstyle[preprint,prd,eqsecnum,aps,epsf]{revtex}

\tightenlines
\begin{document}


    \newcommand{\DSC}{D\hspace{-0.25cm}\slash_{\bot}}
    \newcommand{\DSP}{D\hspace{-0.25cm}\slash_{\|}}
    \newcommand{\DS}{D\hspace{-0.25cm}\slash}
    \newcommand{\DC}{D_{\bot}}
    \newcommand{\DSCX}{D\hspace{-0.20cm}\slash_{\bot}}
    \newcommand{\DSPX}{D\hspace{-0.20cm}\slash_{\|}}
    \newcommand{\DP}{D_{\|}}
    \newcommand{\QV}{Q_v^{+}}
    \newcommand{\QVB}{\bar{Q}_v^{+}}
    \newcommand{\QVP}{Q^{\prime +}_{v^{\prime}} }
    \newcommand{\QVBP}{\bar{Q}^{\prime +}_{v^{\prime}} }
    \newcommand{\QVHZ}{\hat{Q}^{+}_v}
    \newcommand{\QVHZB}{\bar{\hat{Q}}_v{\vspace{-0.3cm}\hspace{-0.2cm}{^{+}} } }
    \newcommand{\QVPHZB}{\bar{\hat{Q}}_{v^{\prime}}{\vspace{-0.3cm}\hspace{-0.2cm}{^{\prime +}}} }
    \newcommand{\QVPHFB}{\bar{\hat{Q}}_{v^{\prime}}{\vspace{-0.3cm}\hspace{-0.2cm}{^{\prime -}} } }
    \newcommand{\QVPHB}{\bar{\hat{Q}}_{v^{\prime}}{\vspace{-0.3cm}\hspace{-0.2cm}{^{\prime}} }   }
    \newcommand{\QVHF}{\hat{Q}^{-}_v}
    \newcommand{\QVHFB}{\bar{\hat{Q}}_v{\vspace{-0.3cm}\hspace{-0.2cm}{^{-}} }}
    \newcommand{\QVH}{\hat{Q}_v}
    \newcommand{\QVHB}{\bar{\hat{Q}}_v}
    \newcommand{\VS}{v\hspace{-0.2cm}\slash}
    \newcommand{\MQ}{m_{Q}}
    \newcommand{\MQP}{m_{Q^{\prime}}}
    \newcommand{\s}{\hat{s}}
    \newcommand{\mk}{\hat{m}_K}
    \newcommand{\mks}{\hat{m}_{K^*}}
    \newcommand{\mb}{\hat{m}_b}
    \newcommand{\mc}{\hat{m}_c}
    \newcommand{\ml}{\hat{m}_l}
    \newcommand{\U}{\hat{u}}
    \newcommand{\mkks}{\hat{m}_{K,K^*}}
   \newcommand{\mbc}{\hat{m}_{b,c}}

\draft
\title{Exclusive B-meson Rare Decays and General Relations of \\ Form Factors
in Effective Field Theory of Heavy Quarks}
\author{M. Zhong$\mbox{}^\dagger$, Y.L. Wu$\mbox{}^\dagger$ and W.Y. Wang$\mbox{}^*$}

\address{$\dagger$ Institute of Theoretical Physics, Academia Sinica,
 Beijing 100080, China \\
 $*$ Department of Physics, Tsinghua University, Beijing 100084, China }
\maketitle

\begin{abstract}
\hspace{0.4cm}
$B$ meson rare decays ($B\to K(K^{*})l\bar l$ and $B\to K^*\gamma$) are analyzed
in the framework of effective field theory of heavy quarks.
The semileptonic and penguin type form factors for these decays are calculated by using the
light cone sum rules method at the leading order of $1/m_Q$ expansion. Four exact relations
between the two types of form factors are obtained at the leading order of $1/m_Q$ expansion.
Of particular, the relations are found to hold for whole momentum transfer region. We also
investigate the validity of the relations resulted from the large energy effective
theory based on the general relations obtained in the present approach.
The branching ratios of the rare decays are presented and their potential importance
for extracting the CKM matrix elements and probing new physics is emphasized.
\end{abstract}


\newpage
\section{Introduction}\label{int}
 $B$ meson rare decays $B\to K^*\gamma$ and $B\to K(K^*)\bar ll$ are induced by transitions $b\to s\gamma$
and $b\to s\bar ll$ via penguin loop diagrams in the quark level, which are usually called
flavor-changing-neutral-current (FCNC) processes. In the
Standard Model (SM), $B$ meson rare decays may provide a quantitative way to determine
the CKM matrix elements $V_{td},V_{ts}$ and $V_{tb}$. In the SM, FCNC transitions are
forbidden at tree level. They can be induced only starting at 1-loop order, which makes their
rates for decaying be sensitive to probe new physics. For these reasons, they have long been
hot subjects in both experimental and theoretical studies.

 Nevertheless, one remains facing difficulties in studying exclusive $B$ meson rare decays due to the requirment of
explicit calculations for the relevant form factors which involve in long distance ingredients
that can not be calculated via QCD perturbative theory. Some reasonable nonperturbative methods, such as QCD sum
rules, lattice simulations and phenomenological models, have been developped to estimate
the long distance effects.

 The rare decays have been studied by using light cone sum rules (LCSR) in full QCD theory \cite{abhh,akos,aos,saf}.
As B meson can be treated as a heavy meson containing a single heavy quark,
it is of interest to apply for the heavy quark effective field theory (HQEFT)
to deal with the B meson rare decays. In this paper, we shall use the framework and normalization
derived in Refs.\cite{ylw,wwy,yww,ww}, which has been applied in Refs.\cite{bpi,brho} to
investigate the exclusive semileptionic $B$ decays into $\pi$ and $\rho$ by using the LCSR method
 and obtained quite reasonable results.  A more general study on exclusive semileptonic decays
of heavy to light mesons within the framework of HQEFT has recently been carried out in \cite{wwz}.
Note that as we only keep to the leading order contributions
in the expansion of $1/m_Q$, the results and conclusions are actually independent of any framework
of effective field theory of heavy quarks.

 We shall focus on in this paper the LCSR calculations of the form factors for the exclusive $B$ meson rare
decays $B\to K^*\gamma$ and $B\to K(K^*)\bar ll$ within the framework of HQEFT.
In section \ref{formulation}, we first present the hadronic matrix elements in the framework
of HQEFT and derive a set of formulae for the relevant form factors. In section \ref{formulation two},
LCSR approach is applied to the relevant correlator functions in HQEFT.
We then obtain four interesting relations among semileptonic type form factors and penguin type ones.
Of interest, these relations are found to hold in whole momentum transfer region in the infinite
mass limit of heavy quark or at the leading order of $1/m_Q$ expansion in HQEFT.
Obviously, at zero recoil of the final light meson, these relations recover
the so-called Isgur-Wise relations\cite{iw}.
In early time, the Isgur-Wise relations were conjectured to be also valid in the region of
large recoil in ref.\cite{BD}. Late on, these relations were really shown in the quark
model\cite{stech,soares} to hold at large recoil. In particular, one of the relations concerning
the form factor in radiative decays was shown to hold in the whole momentum transfer
by using QCD sum rule approach\cite{ABS}. Numerical analysis of form factors is
presented in section \ref{result}. Recently, one developed the so-called large energy
effective theory (LEET) in which more relations were obtained near large recoil
due to additional symmetries in large energy limit\cite{cyopr,efg}, while some of the relations were
found to be broken down by QCD corrections\cite{mbtf}. Since all relations in Ref.\cite{cyopr} were put forward following from 
the combination of heavy quark effective theory (HQET) and LEET and moreover LEET is compatible with LCSR,
our present method provides a reliable and important way to check the validity of LEET relations.  So a detailed discussion and
comparison will be presented on the basis of our HQEFT calaulation in section \ref{leet}. In section \ref{branch ratios},
we give the relevant branching ratios for the B meson rare decays $B\to K^*\gamma$
and $B\to K(K^*)\bar ll$. A brief summary is outlined in section \ref{summary}.

\section{General description of matrix elements in HQEFT}\label{formulation}

The transition matrix elements responsible for the $B$ meson rare decays $B\to K^*\gamma$ and $B\to K(K^*)\bar ll$
may be grouped into two types: semileptonic and penguin ones. The semileptonic ones are defined as
\begin{equation}
\langle K(p)|\bar s\gamma^\mu b|B(p+q) \rangle =2f_{+}(q^2) p^\mu+(f_{+}(q^2)+f_{-}(q^2)) q^\mu
\end{equation}
for $B$ to $K$ decays and
\begin{eqnarray}
&&\langle K^*(p,\epsilon^*)|\bar s\gamma^\mu (1-\gamma^5) b|B(p+q) \rangle =-i(m_B+m_{K^*})A_1(q^2)
\epsilon^{*\mu} \nonumber\\
&&\hspace{2cm}+i \frac{A_2(q^2)}{m_B+m_{K^*}} (\epsilon^{*}\cdot (p+q))(2p+q)^\mu
  +i\frac{A_3(q^2)}{m_B+m_{K^*}} (\epsilon^* \cdot (p+q)) q^\mu \nonumber\\
&&\hspace{2cm} + \frac{2 V(q^2)}{m_B+m_{K^*}} \epsilon^{\mu \alpha \beta \gamma}\epsilon^*_\alpha (p+q)_\beta p_\gamma
\end{eqnarray}
for $B$ to $K^*$ decays. In this paper, we take $\epsilon_{0123} = 1$ and
$\gamma^5=\gamma_5=i\gamma^0\gamma^1\gamma^2\gamma^3$.

For convenience, we may define a form factor $A_0(q^2)$ as
\begin{eqnarray}
A_3(q^2)&=&\frac{2(m_B+m_{K^*})m_{K^*}}{q^2}(\bar A_3(q^2)-A_0(q^2)) \nonumber\\
\bar A_3(q^2)&=&\frac{(m_B+m_{K^*})A_1(q^2)-(m_B-m_{K^*})A_2(q^2)}{2m_{K^*}}\nonumber\\
A_0(0)&=&\bar A_3(0)
\end{eqnarray}
$A_0(q^2)$ will directly enter into contributions to the relevant branching ratios.

The penguin matrix elements can be written as
\begin{equation}
\langle K(p)|\bar s\sigma^{\mu \nu}q_{\nu}(1+\gamma^5)b|B(p+q) \rangle =i\frac{f_T(q^2)}{m_B+m_K}
 \{q^2(2p+q)^\mu -(m^2_B-m^2_K)q^\mu \}
\end{equation}
for $B$ to $K$ decays and
\begin{eqnarray}
&&\langle K^*(p,\epsilon^*)|\bar s\sigma^{\mu \nu}q_{\nu}(1+\gamma^5) b|B(p+q) \rangle =-i\epsilon^{\mu
\alpha \beta \gamma} \epsilon^*_\alpha (p+q)_\beta p_\gamma2T_1(q^2) \nonumber\\
&&\hspace{2cm}+T_2(q^2) \{ (m^2_B-m^2_{K^*})
\epsilon^{*\mu}-(\epsilon^{*}\cdot (p+q))(2p+q)^\mu \} \nonumber\\
&&\hspace{2cm}+T_3(q^2)(\epsilon^* \cdot (p+q))\{ q^\mu-\frac{q^2}{m^2_B-m^2_{K^*}}(2p+q)^\mu \}
\end{eqnarray}
for $B$ to $K^*$ decays.

In the above definitions, $p$ is the momentum of the light meson $K$ or $K^*$. $\epsilon^*$
is the polarization vector of $K^*$ meson, and $q$ is
the momentum transfer. $f_{\pm}$ and $f_T$ are the $B$ to $K$ semileptonic and penguin transition form
factors respectively. $A_i$(i=0,1,2), $V$ and $T_i$(i=1,2,3) are the corresponding ones for $B$ to $K^*$
transitions.

 To be convenient for making Borel transformation which helps to suppress the contributions
 from the possible higher states of bottom mesons, we may change (2.4) and (2.5)
 into the following forms
\begin{equation}
\langle K(p)|\bar s\sigma^{\mu \nu}p_{\nu}(1+\gamma^5)b|B(p+q) \rangle =i\frac{f_T(q^2)}{m_B+m_K}
 \{ (q\cdot p)(2p+q)^\mu -((2p+q)\cdot p)q^\mu \}
\end{equation}
\begin{eqnarray}
&&\langle K^*(p,\epsilon^*)|\bar s\sigma^{\mu \nu}p_{\nu}(1+\gamma^5) b|B(p+q) \rangle =\nonumber\\
&&\hspace{2cm}-i\epsilon^{\mu \alpha \beta \gamma} \epsilon^*_\alpha (p+q)_\beta p_\gamma
\{ \frac{-m^2_B+m^2_{K^*}+q^2}{q^2}T_1(q^2)+\frac{m^2_B-m^2_{K^*}}{q^2}T_2(q^2) \} \nonumber\\
&&\hspace{2cm}+\epsilon^{*\mu}\{ (q\cdot p)\frac{m^2_B-m^2_{K^*}}{q^2}T_2(q^2)
-[(q\cdot p)\frac{m^2_B-m^2_{K^*}}{q^2}-(2p+q)\cdot p ]T_1(q^2) \} \nonumber\\
&&\hspace{2cm}+q^\mu \frac{(\epsilon^* \cdot (p+q))((2p+q)\cdot p)}{m^2_B-m^2_{K^*}}\{ T_3(q^2)
+\frac{m^2_B-m^2_{K^*}}{q^2}(T_2(q^2)-T_1(q^2))\} \nonumber\\
&&\hspace{2cm}-(2p+q)^\mu \frac{(\epsilon^* \cdot (p+q))(q\cdot p)}{m^2_B-m^2_{K^*}}\{T_3(q^2)
+\frac{m^2_B-m^2_{K^*}}{q^2}(T_2(q^2)-T_1(q^2))\}
\end{eqnarray}

 When applying for the HQEFT to evaluate the matrix elements, they can be expanded into the powers of $1/m_Q$ and
also be simply expressed by a set of heavy spin-flavor independent universal wave functions
\cite{wwy,ww,bpi,brho}. It is convenient to adopt the following normalization which
relates matrix elements in full QCD with the ones in HQEFT \cite{wwy}
\begin{eqnarray}
\frac{1}{\sqrt{m_B} }\langle \kappa |\bar s \Gamma b|B \rangle =
\frac{1}{\sqrt{\bar {\Lambda}_B}}
\{ \langle \kappa |\bar s \Gamma b_v|B_v \rangle +O(1/m_b) \}
\end{eqnarray}
where $\kappa$ represents $K(p)$ or $K^*(p,\epsilon^*)$. The notation $b_v$ is the effective bottom quark field. And
$\bar\Lambda_B=m_B-m_b $ is the binding energy.
From heavy quark symmetry, one can obtain the following relations \cite{bpi,brho,kms,gzmy,hly}
\begin{eqnarray}
\langle K(p)|\bar s \Gamma b_v|B_v \rangle &=&-Tr[k(v, p)\Gamma {\cal M}_v]\\
\langle K^*(p,\epsilon^*)|\bar s \Gamma b_v|B_v \rangle &=&-i \mbox{Tr}[\Omega(v, p)\Gamma {\cal M}_v]
\end{eqnarray}
with
\begin{eqnarray}
  k(v, p)&=&\gamma^5 [A(v\cdot p,\mu)+ {\hat{p}\hspace{-0.2cm}\slash}
B(v\cdot p,\mu)] \\
 \Omega(v, p)&=&L_1(v\cdot p) {\epsilon\hspace{-0.2cm}\slash}^*
   +L_2( v\cdot p)(v\cdot \epsilon^*) +[L_3(v\cdot p)
   {\epsilon\hspace{-0.2cm}\slash}^* +L_4(v\cdot p) (v\cdot \epsilon^* )]{\hat{p}\hspace{-0.2cm}\slash}
\end{eqnarray}
and\cite{wwy}
\begin{eqnarray}
{\hat{p}}^\mu &=& \frac{p^\mu}{v\cdot p}  \\
{\cal M}_v &=& -\sqrt{\bar \Lambda}\frac{1+v\hspace{-0.2cm}\slash}{2} \gamma ^5
\end{eqnarray}
 Where $A$, $B$ and $L_i(i=1,2,3,4)$ are the leading order wave functions characterizing the heavy to light
transition matrix elements in the effective field theory. ${\cal M}_v$ is the spin wave function
associated with the heavy meson state. The vector $v^\mu$ is the four-velocity of $B$ meson
satisfying $v^2=1$, and $\bar{\Lambda}$ is the heavy flavor independent binding energy

$$ \bar\Lambda= \lim_{m_Q\to \infty} \bar\Lambda_B $$
which reflects only the effects arising from the light degrees of  freedom in the heavy $B$ meson.

 With eqs.(2.1-2.14), one arrives at the following expressions for the form factors
\begin{eqnarray}
 f_{\pm}(q^2)&=&\frac{1}{m_B} \sqrt{ \frac{m_B \bar\Lambda}
 {\bar{\Lambda}_B } }
    \{ A(v\cdot p)\pm B(v\cdot p) \frac{m_B}{v\cdot p}  \} +\cdots \\
 f_T(q^2)&=&\frac{m_B+m_K}{m_B}\sqrt{\frac{m_B \bar\Lambda}{\bar{\Lambda}_B}}
    \frac{B^{\prime}(v\cdot p)}{v\cdot p} +\cdots
 \\
A_1(q^2)&=&\frac{2}{m_B+m_{k^*}} \sqrt{\frac{m_B \bar\Lambda}{\bar\Lambda_B}}
    \{ L_1(v\cdot p)+L_3(v\cdot p) \}  +\cdots   \\
A_2(q^2)&=&2 (m_B+m_{K^*}) \sqrt{\frac{m_B \bar\Lambda}{\bar\Lambda_B}}
    \{\frac{L_2(v\cdot p)}{2 m^2_B}
     +\frac{L_3(v\cdot p)-L_4(v\cdot p)}{2m_B (v\cdot p)} \}  +\cdots   \\
A_3(q^2)&=&2 (m_B+m_{K^*}) \sqrt{\frac{m_B \bar\Lambda}{\bar\Lambda_B}}
    \{ \frac{L_2(v\cdot p)}{2 m^2_B}
     -\frac{L_3(v\cdot p)
    -L_4(v\cdot p)}{2m_B (v\cdot p)} \}   + \cdots  \\
V(q^2)&=&\sqrt{\frac{m_B \bar\Lambda}{\bar\Lambda_B}}
    \frac{m_B+m_{K^*}}{m_B (v\cdot p) } L_3(v\cdot p) +\cdots \\
T_1(q^2)&=&\sqrt{\frac{m_B \bar\Lambda}{\bar\Lambda_B}}\{\frac{L^{\prime}_1(v\cdot p)}{m_B}+
   \frac{L^{\prime}_3(v\cdot p)}{v\cdot p} \}+\cdots   \\
T_2(q^2)&=&2 \sqrt{\frac{m_B \bar\Lambda}{\bar\Lambda_B}} \frac{1}{m^2_B-m^2_{K^*}}
    \{(m_B-v\cdot p)L^{\prime}_1(v\cdot p)
   +\frac{m_B v\cdot p-m^2_{K^*}}{v\cdot p}
     L^{\prime}_3(v\cdot p) \}+\cdots   \\
T_3(q^2)&=&\sqrt{\frac{m_B \bar\Lambda}{\bar{\Lambda}_B}}\{-\frac{L^{\prime}_1(v\cdot p)}{m_B}+
   \frac{L^{\prime}_3(v\cdot p)}{v\cdot p} -\frac{m^2_B -m^2_{K^*}}{m^2_B v\cdot p}
     L^{\prime}_4(v\cdot p) \}+\cdots
\end{eqnarray}
with
\begin{equation}
y\equiv v\cdot p=\frac{m^2_B+m^2_\kappa -q^2}{2m_B}
\end{equation}
denoting the energy of the final light meson.

 In the above formulae the dots denote possible higher order $1/m_Q$ contributions that are neglected in this paper.
$B^{\prime}(v\cdot p)$ and $L^{\prime}_i(v\cdot p)$  are
in general different from $B(v\cdot p)$ and $L_i(v\cdot p) (i=1,2,3,4)$
as they arise from different matrix elements.

\section{Light-cone sum rules in HQEFT} \label{formulation two}

In order to calculate the relevant hadronic matrix elements which contain nonperturbative contributions
and thus make QCD perturbative method lose its power, we shall apply for LCSR approach.
In LCSR calculation, the relevant correlation functions are expanded near the light cone.
The light cone distribution functions are introduced to describe the nonperturbative effects.
In searching for reasonable and stable results, the quark-hadron duality and Borel transformation are
generally adopted (for a detailed review, one may find in Refs. \cite{vai,ar,pvmisu}).

 The theoretical calculations can often show a simpler process
in the framework of HQEFT than in QCD, which can explicitly be seen in Refs.\cite{bpi,brho} where
the semileptonic form factors for $B\rightarrow \pi$ and $\rho$ have been evaluated.
We may directly adopt the formulae in \cite{bpi,brho,wwz}
to $B\rightarrow K$ transitions by simply changing the relevant quantities corresponding to the $K$ meson

\begin{eqnarray}
 A(y)&=&-\frac{f_K}{4Fy} \int^{s_0}_{0} ds e^{\frac{ 2\bar\Lambda_B-s}{T}}
    \left[\frac{1}{y} \frac{\partial}{\partial u}g_2(u)-{\mu_K} \phi_p(u)
     -\frac{\mu_K}{6}\frac{\partial}{\partial u}\phi_\sigma(u)\right]_{u=1-\frac{s}{2y}} \\
 B(y)&=&-\frac{f_K}{4F} \int^{s_0}_{0} ds e^{\frac{ 2\bar\Lambda_B-s}{T}}
    \left[-\phi_K (u)+\frac{1}{y^2}\frac{\partial^2}{\partial u^2}g_1(u)
    -\frac{1}{y^2} \frac{\partial}{\partial u} g_2(u)
    +\frac{\mu_K}{6y}\frac{\partial}{\partial u} \phi_\sigma(u)\right]_{u=1-\frac{s}{2y}}
\end{eqnarray}

As for the $B$ to vector Kaon meson decays, we may just adopt the definitions of $K^*$ meson
distribution functions given in \cite{pvesrb}.
\begin{eqnarray}
\langle K^*(p,\epsilon^*)|\bar{s}(-x) \sigma_{\mu\nu} d(x)|0 \rangle &=&-i f^T_{K^*}
  [(\epsilon^*_\mu p_\nu - \epsilon^*_\nu p_\mu)
\int^1_0 du e^{-i\xi p\cdot x} (\phi_\bot (u)+\frac{m^2_{K^*}x^2}{4}A_T(u)) \nonumber\\
&+&(p_\mu x_\nu-p_\nu x_\mu)\frac{\epsilon^*\cdot x}{(p\cdot x)^2}m^2_{K^*}\int^1_0 du e^{-i\xi p\cdot x}B_T(u)
+\frac{1}{2}(\epsilon^*_\mu x_\nu \nonumber\\
&-& \epsilon^*_\nu x_\mu)\frac{m^2_{K^*}}{p\cdot x}
\int^1_0 du e^{-i\xi p\cdot x}C_T(u)]   \\
\langle K^*(p,\epsilon^*)|\bar{s}(-x) \gamma_\mu d(x)|0 \rangle &=& f_{K^*} m_{K^*}[ p_\mu
  \frac{\epsilon^*\cdot x}{p\cdot x} \int^1_0 du e^{-i\xi p\cdot x}( \phi_{\|}(u)+\frac{m^2_{K^*}x^2}{4}A(u)) \nonumber\\
  &+& (\epsilon^*_\mu-p_\mu \frac{\epsilon^* \cdot x}{p\cdot x})
  \int^1_0 du e^{-i\xi p\cdot x} g^{(v)}_\bot (u) \nonumber\\
  &-&\frac{1}{2}x_\mu \frac{\epsilon^*\cdot x}{(p\cdot x)^2}m^2_{K^*}\int^1_0 du e^{-i\xi p\cdot x}C(u)] \\
\langle K^*(p,\epsilon^*)|\bar{s}(-x) \gamma_\mu \gamma_5 d(x)|0 \rangle &=&\frac{1}{2} (f_{K^*}-f^T_{K^*}
\frac{m_s+m_d}{m_{K^*}}) m_{K^*} \epsilon_{\mu\nu\alpha\beta} \epsilon^{*\nu} p^\alpha x^\beta
  \int^1_0 du e^{-i\xi p\cdot x} g^{(a)}_\bot (u)  \\
\langle K^*(p,\epsilon^*)|\bar{s}(-x)d(x)|0 \rangle &=& i(f^T_{K^*}-f_{K^*}\frac{m_s+m_d}{m_{K^*}})
 m^2_{K^*} ( \epsilon^*\cdot x)\int^1_0 du e^{-i\xi p\cdot x} h^{(s)}_{\|}(u)
\end{eqnarray}
with $\xi=2u-1$. In the above definitions $\phi_{\bot}(u)$ and $\phi_{\|}(u)$ give the
leading twist two distributions in the fraction of total momentum carried by quarks in the transversely and longitudinally
polarized meson respectively. The functions $g^{(v)}_\bot (u)$ and $g^{(a)}_{\|}(u)$, which describe transverse
polarizations of quarks in the longitudinally polarized meson, provide contributions of twist three. $h^{(s)}_{\|}$ is also
twist three function. $A(u)$, $A_T(u)$ and $C(u)$ are twist four, while $B_T(u)$ and $C_T(u)$ are mixing twist functions.
In this paper, we only consider distribution functions from two-particle contributions.

Applying the same calculation steps as in \cite{brho}, we arrive at the following results
\begin{eqnarray}
L_1(y)&=& \frac{1}{4F} e^{2\bar \Lambda_B/T}
   \int^{s_0}_0 ds e^{-s/T} \frac{1}{y} \left[f_{K^*} m_{K^*}g^{(v)}_\bot(u)\right.\nonumber\\
   & &\left.+\frac{1}{4}(f_{K^*}-f^T_{K^*}\frac{m_s+m_d}{m_{K^*}})m_{K^*}\frac{\partial}{\partial u} g^{(a)}_\bot (u)
    +\frac{f^T_{K^*}m^2_{K^*}}{2y}C_T(u) \right]_{u=1-\frac{s}{2y}}  \\
L_2(y)&=& \frac{1}{4F} e^{2\bar\Lambda_B/T}
   \int^{s_0}_0 ds e^{-s/T}
   \left[\frac{f^T_{K^*}m^2_{K^*}}{y^2}(\frac{1}{2}C_T(u)+B_T(u)+\frac{1}{2}\frac{\partial}{\partial u}h^{(s)}_\|(u))\right. \nonumber\\
   & &\left.+f_{K^*}m^2_{K^*}(\frac{m_{K^*}}{2y^3}C(u)-\frac{m_s+m_d}{2y^2m_{K^*}}
   \frac{\partial}{\partial u}h^{(s)}_\|(u))\right]_{u=1-\frac{s}{2y}}  \\
L_3(y)&=& \frac{1}{4F} e^{2\bar\Lambda_B/T}
   \int^{s_0}_0 ds e^{-s/T}
   \left[-\frac{1}{4y} (f_{K^*}-f^T_{K^*}\frac{m_s+m_d}{m_{K^*}}) m_{K^*} (\frac{\partial}{\partial u}
   g^{(a)}_\bot(u) )\right. \nonumber\\
   & &\left.+f^T_{K^*} (\phi_\bot(u)-\frac{m^2_{K^*}}{16y^2}
   \frac{\partial ^2}{\partial u^2}A_T(u))\right]_{u=1-\frac{s}{2y}}  \\
L_4(y)&=& \frac{1}{4F} e^{2\bar\Lambda_B/T}
   \int^{s_0}_0 ds e^{-s/T} \frac{1}{y} \nonumber\\
   & &\left[f_{K^*} m_{K^*} ( \phi_{\|}(u)-g^{(v)}_\bot(u)
  -\frac{1}{4} \frac{\partial}{\partial u} g^{(a)}_\bot (u)-\frac{m^2_{K^*}}{16y^2}
   \frac{\partial ^2}{\partial u^2}A(u))\right. \nonumber\\
   & &\left.+\frac{f^T_{K^*}m^2_{K^*}}{y}B_T(u)
   +\frac{1}{4}f^T_{K^*}(m_s+m_d)\frac{\partial}{\partial u}g^{(a)}_\bot (u)\right]_{u=1-\frac{s}{2y}}
\end{eqnarray}

  For the penguin matrix elements, we begin with considering the following correlations
\begin{eqnarray}
P_T^\mu(p,q)&=&i\int d^4x e^{iq\cdot x}\langle K(p)|T\{\bar s(x)\sigma^{\mu\nu}p_\nu(1+\gamma^5)b(x),
\bar b(0)i\gamma^5d(0)\}|0\rangle \\
V_T^\mu(p,q)&=&i\int d^4x e^{-ip_B\cdot x}\langle K^*(p,\epsilon^*)|T\{\bar s(0)\sigma^{\mu\nu}p_\nu(1+\gamma^5)b(0),
\bar b(x)i\gamma^5d(x)\}|0\rangle
\end{eqnarray}
where $p_B=p+q$.

  Making the procedures similar to Refs.\cite{bpi,brho}, one can easily derive the explicit
expressions for $B^{\prime}(y)$ and $L^{\prime}_i(y)$, which are exactly the same as $B(y)$ and $L_i(y)$ in (3.2) and (3.7-3.10)
at the leading order of $1/m_Q$ expansion. From (2.15) to (2.23), we obtain the following four interesting relations
among the semileptonic type and penguin type form factors.
\begin{eqnarray}
f_T(q^2) & = & \frac{m_B+m_K}{2m_B}(f_+(q^2)-f_-(q^2))   \\
T_1(q^2)&=&\frac{m^2_B-m^2_{K^*}+q^2}{2m_B}\frac{V(q^2)}{m_B+m_{K^*}}+\frac{m_B+m_{K^*}}{2m_B}A_1(q^2)   \\
T_2(q^2)&=&\frac{2}{m^2_B-m^2_{K^*}}[\frac{(m_B-y)(m_B+m_{K^*})}{2}A_1(q^2)+\frac{m_B(y^2-m^2_{K^*})}{m_B+m_{K^*}}V(q^2)]  \\
T_3(q^2)&=&\frac{m^2_B-m^2_{K^*}}{m_Bq^2}m_{K^*}A_0(q^2)-\frac{m^2_B-m^2_{K^*}+q^2}{2m_Bq^2}
           [(m_B+m_{K^*})A_1(q^2)-(m_B-m_{K^*})A_2(q^2)] \nonumber\\
  & &+\frac{m^2_B+3m^2_{K^*}-q^2}{2m_B(m_B+m_{K^*})}V(q^2)
\end{eqnarray}
  Note that these relations hold in the whole momentum transfer region.
  In fact, they can simply be obtained from eqs.(2.9)
 and (2.10) by assuming that the light meson wave functions
 $k(v, p)$ and $\Omega(v, p)$ are universal for different types
 of the currents, namely they are independent of the choice of
 $\Gamma$ in the currents. The second relation was actually
 observed in ref.\cite{ABS} by using QCD sum rule approach.

  In the heavy quark limit near zero recoil point ($q^2\rightarrow q^2_{max}$),
  the above relations were known as Isgur-Wise relations\cite{iw}.
In Refs. \cite{pvesrb,pab}, some detailed discussions were made to
explore the Isgur-Wise relations in both small and large recoil conditions by using full LCSR method,
where it was shown that the Isgur-Wise relations are satisfied very well at $q^2\rightarrow 0$ (large recoil)
and hold with about $80\%$ accuracy at large $q^2$.
In Ref.\cite{cfss}, conclusions were made that the
Isgur-Wise relations are valid up to $70\%$ in the whole $q^2$ region by applying for the three point QCD sum rules method.
 In the quark model, the authors of Refs.\cite{stech,soares}
concluded that Isgur-Wise relations also hold at large recoil.

  In here we have further confirmed the existed analyzes in the literature and
  arrived at a general conclusion that the relations (3.13-3.16)
  hold for whole allowed region of momentum transfer $q^2$ at the
 leading order of $1/m_Q$ expansion, which has nothing to do with the energy of the final light meson.
 This may provide a more general proof in support of
 the hypothesis made in the early time by the authors in Ref. \cite{BD}.
 As a consequence, it becomes remarkable that one
can directly read off the penguin form factors $f_T$ and $T_i(i=1,2,3)$  from
the relations in eqs. (3.13-3.16) without the tedious calculations from LCSR to Borel transformation.
In Ref.\cite{wwz}, we have shown that when applying the HQEFT to heavy to light semileptonic decays,
heavy quark expansion and heavy quark symmetry enable us to relate various decay channels, consequently, the theoretical
analysis is much simplified and the number of independent functions is greatly abated though the number of independent
functions in a single decay channel does not decrease. Here with the four exact relations, the number of independent variables
among the form factors $f_{\pm}$ and $f_T$ is straightforwardly reduced to two. The number of independent variables in
 the rare decay $B\to K^*l\bar l$ is then reduced from seven form factors to four form factors.
 One sees that for the $B$ meson rare decays the heavy quark expansion exhibits
 its more powerful advantages, the number of independent functions
 is found to be largely reduced even in a single decay channel.

\section{Numerical analysis and results for form factors}\label{result}

  The light cone wave functions play an important role for a precise calculation of form factors.
They have been studied by several groups. We shall use the results
given in \cite{ar,arsco,vvar,vi} for $K$ and the ones in
\cite{pvesrb} for $K^*$ meson in our following analyses. The
asymptotic form and the scale dependence of them are taken from
perturbative QCD calculations given in \cite{va,bf}.

 We shall first present all two-particle light cone amplitudes of Kaon appearing in (3.1) and (3.2)
\begin{eqnarray}
\phi_K(u,\mu)&=&6u(1-u)\left[1+a_1(\mu)(3(2u-1))+a_2(\mu)(\frac{15}{2}(2u-1)^2-\frac{3}{2})\right. \nonumber\\
              & &\left.+a_3(\mu)(\frac{35}{2}(2u-1)^3-\frac{15}{2}(2u-1))+
                        a_4(\mu)\frac{15}{8}(21(2u-1)^4-14(2u-1)^2+1)\right] \nonumber\\
\phi_p(u,\mu)&=& 1+\frac{1}{2}B_2(\mu)[3(2u-1)^2-1]+\frac{1}{8}B_4(\mu) [35 (2u-1)^4-30 (2u-1)^2+3] \nonumber\\
\phi_\sigma(u,\mu)&=& 6u(1-u)\{ 1+\frac{3}{2}C_2(\mu) [5(2u-1)^2-1]+\frac{15}{8}C_4(\mu)
  [21 (2u-1)^4-14(2u-1)^2+1]\}  \nonumber\\
g_1(u,\mu)&=& \frac{5}{2}\delta^2(\mu) u^2 (1-u)^2+\frac{1}{2}\epsilon(\mu) \delta^2(\mu)
  \left[u(1-u)(2+13u (1-u)+10u^3 (\log u)(2-3u+\frac{6}{5}u^2) \right.\nonumber\\
  & &\left.+10(1-u)^3 (\log(1-u))(2-3(1-u)+\frac{6}{5}(1-u)^2))\right] \nonumber\\
g_2(u,\mu)&=&\frac{10}{3} \delta^2(\mu) u(1-u)(2u-1)
\end{eqnarray}
For the relevant parameters, we will take the following values in our numerical calculations
\begin{eqnarray}
&a_1(\mu_b)=0.15, \;\; a_2(\mu_b)=0.16, \;\;
a_3(\mu_b)=0.05, \;\; a_4(\mu_b)=0.06,   \nonumber\\
&B_2(\mu_b)=0.29, \;\; B_4(\mu_b)=0.58, \;\;
C_2(\mu_b)=0.059, \;\; C_4(\mu_b)=0.034, \nonumber\\
&\delta^2(\mu_b)=0.17 \mbox{GeV}^2, \;\;
\epsilon(\mu_b)=0.36
\end{eqnarray}
where we choose $\mu_b=\sqrt{m^2_B-m^2_b}\approx 2.4$GeV which is the appropriate scale characterizing the typical
virtuality of the b quark. The SU(3) flavor violation effects of Kaon has been taken into account in $\phi_K(u,\mu)$ with
 non-vanishing coefficients $a_1(\mu_b)$ and $a_3(\mu_b)$. The SU(3) flavor breaking effects in the higher twist amplitudes
 are neglected in this paper as they only have a small contribution \cite{pb}.

For the light vector $K^*$ meson wave functions, we use the expressions given in \cite{pvesrb}.
The light meson SU(3) flavor breaking effects are taken into account in the leading twist distributions
and partially in the twist three, but ignored in twist four.

Other parameters needed are listed in the following :
\begin{eqnarray}
&m_B=5.28 \mbox{GeV}, \;\; m_b=4.75 \mbox{GeV}, \;\;
m_K=0.49 \mbox{GeV}, \;\; m_{K^*}=0.89 \mbox{GeV}, \nonumber\\
&\bar\Lambda_B=0.53 \mbox{GeV}, \;\; \bar{\Lambda}=(0.53\pm 0.08) \mbox{GeV},\;\;
F=(0.30\pm 0.06)\mbox{GeV}^{3/2},\nonumber\\
&f_K=0.16 \mbox{GeV}, \;\; f_{K^{*}}=(226 \pm 28) \mbox{MeV}, \;\;
f^{\bot}_{K^{*}}(\mu_b)=(175 \pm 9) \mbox{MeV} ,\nonumber\\
&\mu_K(\mu_b)=\frac{m^2_K}{m_s+m_{u,d}}\approx 2.02 \mbox{GeV}
\end{eqnarray}

As for the parameters $s_0$ and $T$, according  to LCSR criterion that the contributions from both the higher
states and higher twist four distribution functions should not be larger than 30\%. Thus we
choose the  region for $T$ to be $ 1\mbox{GeV}<T<3 \mbox{GeV}$. In this region the curves of
$f_{\pm}$ and $f_T$ for $B$ to $K$ transitions
and $A_i(i=0,1,2)$, $V$ and $T_i(i=1,2,3)$ for $B$ to $K^*$ transitions become most stable at the threshold energy
$s_0=2.1-2.7 (2.4\pm 0.3)$GeV and $s_0=1.8-2.4 (2.1\pm 0.3)$GeV respectively, which may be seen explicitly from Fig.1 to Fig.10.

 It is known that  the light cone expansion and the sum rule method may be broken down at large momentum transfer
(practically as $q^2$ approaches near half of $m^2_b$)\cite{ar}, which may be seen explicitly from Fig.11-20.
Where the curves of form factors calculated from LCSR likely become unstable at large $q^2$ region.
Thus for the behavior of the form factors in the whole kinematically accessible
region, we may use the following parametrization for the transfer momentum dependence of
the form factors
\begin{equation}
F(q^2)=\frac{F(0)}{1-a_F q^2/m^2_B+b_F (q^2/m^2_B)^2}
\end{equation}
where $F(q^2)$ can be any of the form factors $f_+$, $f_-$, $f_T$, $ A_i(i=0,1,2)$, $V$ and $T_i(i=1,2,3)$.
We directly use LCSR predictions to fit the parameters. This is because $K^{\ast}$ and $K$ mesons contain
 a relative heavy strange quark (in comparison with the $u$ and $d$ quarks), so that they are comparatively heavy and
 shorten the kinematically allowed ranges of $B$ to $K^{\ast}$ and $K$ decays in comparison with
 the ranges for $B$ to $\pi$ decays. Therefore the sum rules are expected to yield
 reasonable values for most allowed region of $q^2$ in the $K^{\ast}$ and $K$ meson cases.

 With the above analyses, we are then able to fix the three parameters for each
 form factor. We plot in Fig.11-20 the form factors as functions of $q^2$ with different threshold energy $s_0=2.1, 2.4, 2.7$GeV
 for $B$ to $K$ transitions and $s_0=1.8, 2.1, 2.4$GeV for $B$ to $K^*$ transitions at $T=2.0$GeV.
 Our numerical results of form factors at $q^2=0$ are given in Table 1,
 where the uncertainties of the form factors
arise from the uncertainties of the threshold energy $s_0$, the
Borel parameter $T$ and the parameters in (4.3). In our values,
the uncertainties coming from $s_0$ are about 10\%, from $T$ and
the parameters in (4.3) are around 15\%. The total uncertainties
are up to 25\%.  For comparison, we list in table 1 the numerical
 results obtained from other approaches: QCD LCSR, LEET, quark model (QM), lattice, three point
sum rules (SR) and PQCD calculation.

\begin{center}
\parbox{12cm}{
\small
\baselineskip=1.0pt
Table 1. Values for form factors of $B\to Kl\bar l$,$B\to K^*l\bar l$ and $B\to K^*\gamma$ at $q^2=0$.
}
\end{center}
\begin{center}
\begin{tabular}{|c|c|c|c|c|c|c|c|c|}
\hline \hline
  & present & LCSR\cite{abhh} & LEET\cite{BH} & QM\cite{MS} & lattice\cite{UKQCD} & lattice\cite{ABA}
  & SR\cite{cfss} & PQCD\cite{TAI} \\
\hline
 $f_+(0)$ & 0.454 $^{+0.053}_{-0.075}$ & 0.319 $^{+0.052}_{-0.041}$ & $---$  & 0.36 &$---$&0.30(4)& 0.25 &$---$\\
\hline
 $f_0(0)$ & 0.454 $^{+0.053}_{-0.075}$  & 0.319 $^{+0.052}_{-0.041}$ & $---$  & 0.36 &$---$& 0.30(4) & 0.25&$---$  \\
\hline
 $f_T(0)$ & 0.447 $^{+0.046}_{-0.069}$ & 0.355 $^{+0.016}_{-0.055}$ & $---$  & 0.35 &$---$& 0.29(6)  & 0.14&$---$ \\
\hline
 $A_0(0)$ & 0.468 $^{+0.082}_{-0.112}$   & 0.471 $^{+0.127}_{-0.059}$ & $---$   &  0.45 & 0.32 &$---$& 0.30 & 0.407  \\
\hline
 $A_1(0)$ & 0.350 $^{+0.068}_{-0.089}$  & 0.337 $^{+0.048}_{-0.043}$  & 0.27 $\pm$0.03 & 0.36& 0.28 &$---$& 0.37 & 0.266\\
\hline
 $A_2(0)$ & 0.302 $^{+0.063}_{-0.080}$ & 0.282 $^{+0.038}_{-0.036}$ & $---$  &  0.32 &$---$  &$---$ & 0.40 & 0.202\\
\hline
 $V(0)$ & 0.426 $^{+0.070}_{-0.098}$ & 0.457 $^{+0.091}_{-0.058}$ & 0.36 $\pm$0.04 & 0.44 & 0.38 &$---$& 0.47& 0.355   \\
\hline
 $T_1(0)$ & 0.382 $^{+0.068}_{-0.093}$ & 0.379 $^{+0.058}_{-0.045}$  & 0.31 $\pm$0.02 &  0.39 & 0.32  &$---$& 0.38 & 0.315 \\
\hline
 $T_2(0)$ & 0.382 $^{+0.068}_{-0.093}$ & 0.379 $^{+0.058}_{-0.045}$ & $---$  &  0.39 & 0.32 &$---$& 0.38 & 0.315\\
\hline
 $T_3(0)$ & 0.266 $^{+0.045}_{-0.063}$ & 0.260 $^{+0.035}_{-0.016}$  & $---$  &  0.27 &  &$---$& 1.4 & 0.207\\
\hline \hline
\end{tabular}
\end{center}

\vspace{0.2cm}

\section{On validity of relations in LEET}\label{leet}

 Recently, it was noted in Ref.\cite{cyopr} that by applying
 HQET to the initial heavy meson
 and meanwhile adopting LEET to describe the final light meson,
 more relations were obtained at the leading order of heavy quark mass and large energy expansion.
 Eventually, the total number of independent form factors is reduced to three near the large recoil point.
 While those symmetry relations were shown\cite{mbtf} to be broken down when radiative corrections are considered.
   The contributions of the second order in the
 ratio of the light meson mass to the large recoil energy were taken into account in Ref.\cite{efg}.

  To compare our results with those in Ref.\cite{cyopr} and also to see how much accuracy of the large
  energy effective theory would be, we first introduce the following
  definitions
\begin{eqnarray}
\zeta(m_B,y) & = & \sqrt{\frac{m_B \bar\Lambda}{\bar\Lambda_B}}\frac{1}{y}B(y)  \\
\zeta_A(m_B,y) & = & \sqrt{\frac{m_B \bar\Lambda}{\bar\Lambda_B}}\frac{1}{m_B}A(y)  \\
\zeta_{\parallel}(m_B,y) & = & \sqrt{\frac{m_B \bar\Lambda}{\bar\Lambda_B}}(\frac{1}{m_{K^*}}
                             L_4(y)-\frac{y}{m_Bm_{K^*}}L_2(y)) \\
\zeta_{\perp}(m_B,y) & = & \sqrt{\frac{m_B \bar\Lambda}{\bar\Lambda_B}}\frac{L_3(y)}{y} \\
\zeta_1(m_B,y) & = & \sqrt{\frac{m_B \bar\Lambda}{\bar\Lambda_B}}\frac{L_1(y)}{y}  \\
\zeta_2(m_B,y) & = & \sqrt{\frac{m_B \bar\Lambda}{\bar\Lambda_B}}\frac{y}{m_Bm_{K^*}}L_2(y)
\end{eqnarray}
 With these functions, the form factors in eqs.(2.15-2.23) can be expressed as
\begin{eqnarray}
f_{\pm}(q^2) & = & \zeta_A(m_B,y)\pm \zeta(m_B,y)   \\
f_0(q^2) & \equiv & \frac{q^2}{m^2_B-m^2_{K}}f_-(q^2)+f_+(q^2)   \nonumber\\
         & = & (1-\frac{q^2}{m^2_B-m^2_{K}})\zeta(m_B,y)+(1+ \frac{q^2}{m^2_B-m^2_{K}})\zeta_A(m_B,y) \\
f_T(q^2) & = & (1+\frac{m_{K}}{m_B})\zeta(m_B,y)  \\
A_0(q^2) & = & (1-\frac{m^2_{K^*}}{ym_B})\zeta_{\parallel}(m_B,y)+\frac{m_{K^*}}{m_B}\zeta_{\perp}+
               \frac{y}{m_{K^*}}\zeta_1(m_B,y)-\frac{q^2}{ym_B}\zeta_2(m_B,y)  \\
A_1(q^2) & = & \frac{2y}{m_B+m_{K^*}}(\zeta_{\perp}(m_B,y)+\zeta_1(m_B,y))  \\
A_2(q^2) & = & (1+\frac{m_{K^*}}{m_B})(\zeta_{\perp}(m_B,y)-\frac{m_{K^*}}{y}\zeta_{\parallel}(m_B,y))  \\
V(q^2)   & = & (1+\frac{m_{K^*}}{m_B})\zeta_{\perp}(m_B,y)   \\
T_1(q^2) & = & \zeta_{\perp}(m_B,y)+\frac{y}{m_B}\zeta_1(m_B,y)  \\
T_2(q^2) & = & (1-\frac{q^2}{m^2_B-m^2_{K^*}})\zeta_{\perp}(m_B,y)+(1+ \frac{q^2}{m^2_B-m^2_{K^*}})
               \frac{y}{m_B}\zeta_1(m_B,y)  \\
T_3(q^2) & = & \zeta_{\perp}(m_B,y)-(1-\frac{m^2_{K^*}}{m^2_{m_B}})\frac{m_{K^*}}{y}(\zeta_{\parallel}(m_B,y)+\zeta_2(m_B,y))
               -\frac{y}{m_B}\zeta_1(m_B,y)
\end{eqnarray}

   Comparing the above forms with the ones given by eqs.(104)-(113) in Ref.\cite{cyopr}, it is obvious that
the three functions $\zeta_A(q^2)$, $\zeta_1(q^2)$ and  $\zeta_2(q^2)$ should
vanish, i.e., $\zeta_A(q^2) = \zeta_1(q^2) = \zeta_2(q^2) = 0$, in order to reproduce the
results in LEET.  For a quantitative comparison, we plot in Fig.21
three curves $\zeta_{\parallel}(q^2)$, $\zeta_{\perp}(q^2)$ and $\zeta(q^2)$ as functions of momentum transfer
$q^2$, and in Fig.22 the three ratios $\zeta_2(q^2)/\zeta_{\parallel}(q^2)$,
$\zeta_1(q^2)/\zeta_{\perp}(q^2)$ and $\zeta_A(q^2)/\zeta(q^2)$ as
functions of $q^2$. It is seen that the three functions $\zeta_A(q^2)$,
$\zeta_1(q^2)$
and $\zeta_2(q^2)$ do have  small but sizable contributions relative
to the
three nonzero functions $\zeta_{\parallel}(q^2)$,
 $\zeta_{\perp}(q^2)$ and
$\zeta(q^2)$ in LEET. Numerically, they are
about 20\%, 10\% and 10\% respectively when $q^2 = 0$. 
For the ratios $\zeta_2(q^2)/\zeta_{\parallel}(q^2)$ and $\zeta_A(q^2)/\zeta(q^2)$, 
they are almost independent of the
momentum transfer $q^2$. For the ratio
$\zeta_1(q^2)/\zeta_{\perp}(q^2)$, it ranges from $10\%$ to $40\%$ as $q^2$
increases.
    
    To be more clear, we directly plot in Figs.23-29 the ratios between the
form factors,
i.e., $F_{iLEET}(q^2)/F_i(q^2)$ at center value of $s_0$ with $T=2.0$GeV.
Here $F_{i LEET}(q^2)$ denote the form factors obtained from
 HQET/LEET
relations, namely  $\zeta_A(q^2) = \zeta_1(q^2) =
 \zeta_2(q^2) = 0$, and
$F_i(q^2)$s are the form factors obtained
 in present paper, namely the three
functions $\zeta_A(q^2)$,
 $\zeta_1(q^2)$ and $\zeta_2(q^2)$ are not zero and
given in eq.
(5.2), (5.5) and (5.6). It is seen that, except the ratio
$A_{0 LEET}/A_0 \approx
 0.78$, the other ratios near large
recoil point
($q^2\rightarrow 0$) have the value $F_{i LEET}/F_i = 0.9\sim 1.0$. 
More explicitly speaking $f_{+ LEET}/f_+ \approx f_{T LEET}/f_T \approx0.9$, $A_{1 LEET}/A_1 \approx 0.92$,
$T_{1 LEET} \approx T_{2 LEET}/T_2 \approx 0.96$ and $T_{3 LEET}/T_3 \approx 0.98$.
So we can say the form factors 
$F_{i LEET}$ have the accuracy beter than $90\%$
as comparation with $F_i$
except $A_{0 LEET}$ whose accuracy is only about
$78\%$. This exception
can be explained as that the third term on the R.H.S. of the equation (5.10) contains the
factor $y/m_{K^*}$, which undoubtedly enhances several times the
contribution of $\zeta_1(q^2)$ when $q^2 = 0$. So when $B$
decays to more light final mesons,
say $\rho$, the LEET
relation for $A_0(q^2)$ will work worse.  It
is
particularly noted that ratios $f_{+ LEET}/f_+$,
$T_{1 LEET}/T_1$ and
$T_{3 LEET}/T_3$, which take value among
the area $0.85\sim 0.90$, $0.91\sim 0.95$ and $1.0\sim 1.05$ respectively in the whole $q^2$, are almost independent of $q^2$.
The reason is clear since the ratios $\zeta_A(q^2)/\zeta(q^2)$ in (5.7) and $\zeta_2(q^2)/\zeta_{\parallel}(q^2)$ in (5.16) are
almost independent of the
momentum transfer $q^2$ and the coefficient of the wave
function $\zeta_1(m_B, y)$ is associated with the factors $y/m_B$ in (5.14) and (5.16), which have a tendency to suppress the contribution
of $\zeta_1(m_B, y)$ as $q^2$ increases. While the other ratios, especially the ratio $A_{0 LEET}/A_0$,
get worse results for large $q^2$, say $q^2 \approx 15GeV^2$, which can be considered as far from large
recoil. For instance $A_{0 LEET}$ is only about half of $A_0$ at $q^2 \approx 15GeV^2$. This means that LEET becomes
not appropriate for large $q^2$ region.
The reason lies in that on one hand $\zeta_1(q^2)/\zeta_{\perp}(q^2)$ counts notablly when $q^2$ is large.
On the other hand, in the general form factor formulations (5.8), (5.10) and (5.15), there is a factor $q^2$
to enhance the contributions of $\zeta_A(q^2)$, $\zeta_2(q^2)$ and $\zeta_1(q^2)$ when $q^2$ increases.
In addition, as can be seen from (5.9), (5.12) and (5.13), $f_{T LEET}$, $A_{2 LEET}$ and $V_{LEET}$
are same as the corresponding one obtained by HQEFT in present paper.

 Besides the relations (3.13-3.16), LEET leads to  three additional relations \cite{cyopr}
\begin{eqnarray}
f_T(q^2) & = & (1+\frac{m_{K}}{m_B})f_+(q^2)  \\
A_0(q^2) & = & (1-\frac{m^2_{K^*}}{ym_B})\frac{ym_B}{m_{K^*}(m_B+m_{K^*})}(V(q^2)-A_2(q^2))
+\frac{m_{K^*}(m_B+m_{K^*})}{2ym_B}A_1(q^2)  \\
A_1(q^2) & = & \frac{2ym_B}{(m_B+m_{K^*})^2}V(q^2)
\end{eqnarray}
To check this relations, we plot in  Figs.30-32 the three
form factors $f_T$, $A_0$ and $A_1$ obtained from the direct
calculations (i.e., left-hand side (LHS) ) and from the relations
(i.e., right-hand side (RHS)). The numerical comparison of the LHS
and RHS of the above equations shows that near the large recoil
point ($q^2\rightarrow 0$) the relations (5.17) and (5.19) both hold within 90\%
accuracy, but the relations (5.18) only hold about 80\% accuracy. In particular, the
deviations of LHS and RHS in relations (5.17) and (5.19) do not exceed 20\% through whole $q^2$,
while that of relation (5.18) at best hold with 40\%.

\section{Results for branching ratios}\label{branch ratios}

The relevant branching ratios are able to be calculated with the
form factors given above. The relevant decay width formulae have
the following general forms \cite{abhh}

\begin{eqnarray}
\frac{d\Gamma}{d{\s}}&=&
\frac{G^2_F \alpha ^2 m^5_B}{2^{11} \pi ^5} |V^*_{ts}V_{tb}|^2 {\U} ({\s})\left\{(|A'({\s})|^2+|C'({\s})|^2)
         (\lambda ({\s})-\frac{{\U}({\s}) ^2}{3}) \right.\nonumber\\
         & & \left.+|C'({\s})|^2 4{\ml}^2(2+2{\mk}^2-{\s})+Re(C'({\s})D'({\s})^*)8{\ml} ^2(1-{\mk}^2)+|D'({\s})|^24{\ml}^2{\s}\right\}
\end{eqnarray}
for $B\rightarrow K \bar{l}l$, and
\begin{eqnarray}
\frac{d\Gamma}{d{\s}}&=&\frac{G^2_F \alpha ^2  m^5_B}{2^{11} \pi ^5} |V^*_{ts}V_{tb}|^2 {\U}
                      ({\s})\left\{\frac{|A({\s})|^2}{3}{\s}\lambda ({\s})
         (1+2\frac{{\ml}^2}{{\s}})+|E({\s})|^2{\s}\frac{{\U}({\s})^2}{3} \right.\nonumber \\
         & &+\frac{1}{4{\mks}^2}\left[|B({\s})|^2 (\lambda ({\s})-\frac{{\U}({\s})^2}{3}+
                      8{\mks}^2 ({\s}+2{\ml}^2)) +|F({\s})|^2(\lambda ({\s})
         -\frac{{\U}({\s})^2}{3}+8{\mks}^2 ({\s}-4{\ml}^2))\right] \nonumber \\
         & &+\frac{\lambda ({\s})}{4{\mks}^2}\left[|C({\s})|^2(\lambda ({\s})-\frac{{\U}({\s})^2}{3})
         +|G({\s})|^2(\lambda ({\s})-\frac{{\U}({\s})^2}{3}+4{\ml}^2(2+2{\mks}^2-{\s}))\right]  \nonumber \\
         & &-\frac{1}{2{\mks}^2}\left[Re(B({\s})C({\s})^*) (\lambda ({\s})-\frac{{\U}({\s})^2}{3})(1-{\mks}^2-{\s})\right.\nonumber \\
         & &\left.+Re(F({\s})G({\s})^*) ((\lambda ({\s})-\frac{{\U}({\s})^2}{3})
                     (1-{\mks}^2-{\s})+4{\ml}^2\lambda ({\s}))\right] \nonumber \\
         & &\left.-2\frac{{\ml}^2}{{\mks}^2}\lambda ({\s})\left[Re(F({\s})H({\s})^*)-Re(G({\s})H({\s})^*)(1-{\mks}^2)\right]
                   +\frac{{\ml}^2}{{\mks}^2}{\s}\lambda ({\s})|H({\s})|^2 \right\}
\end{eqnarray}
for $B\rightarrow K^{\ast} \bar{l}l$,  as well as
\begin{equation}
\Gamma =
\frac{G^2_F \alpha m^2_b m^3_B}{32 \pi ^4} |V^*_{ts}V_{tb}|^2 |C^{eff}_7|^2 |T_1(0)|^2 (1-{\mks})^3
\end{equation}
for $B\rightarrow K^{\ast} \gamma$.

 The functions appearing in the above decay width formulae are defined as \cite{abhh,bm}
\begin{eqnarray}
A'({\s}) &=& C^{eff}_9({\s})f_+({\s})+\frac{2{\mb}}{1+{\mk}}C^{eff}_7f_T({\s}),\\
B'({\s}) &=& C^{eff}_9({\s})f_-({\s})-\frac{2{\mb}}{{\s}}(1-{\mk})C^{eff}_7f_T({\s}),\\
C'({\s}) &=& C_{10}f_+({\s}),\\
D'({\s}) &=& C_{10}f_-({\s}),\\
A({\s}) &=& \frac{2}{1+{\mks}}C^{eff}_9({\s})V({\s})+\frac{4{\mb}}{{\s}}C^{eff}_7T_1({\s}),\\
B({\s}) &=& (1+{\mks})[C^{eff}_9({\s})A_1({\s})+\frac{2{\mb}}{{\s}}(1-{\mks})C^{eff}_7T_2({\s})],\\
C({\s}) &=& \frac{1}{1-{\mks}^2}[(1-{\mks})C^{eff}_9({\s})A_2({\s})+2{\mb}C^{eff}_7(T_3({\s})+\frac{1-{\mks}^2}{{\s}}T_2({\s}))],\\
D({\s}) &=& \frac{1}{{\s}}[C^{eff}_9({\s})((1+{\mks})A_1({\s})-(1-{\mks})A_2({\s})-2{\mks}A_0({\s}))-2{\mb}C^{eff}_7T_3({\s})],\\
E({\s}) &=& \frac{2}{1+{\mks}}C_{10}V({\s}),\\
F({\s}) &=& (1+{\mks})C_{10}A_1({\s}),\\
G({\s}) &=& \frac{1}{1+{\mks}}C_{10}A_2({\s}),\\
H({\s}) &=& \frac{1}{{\s}}C_{10}({\s})[(1+{\mks})A_1({\s})-(1-{\mks})A_2({\s})-2{\mks}A_0({\s})],\\
\lambda ({\s}) &=& 1+{\mkks}^4+{\s}^2-2{\s}-2{\mkks}^2(1+{\s}),\\
{\U}({\s}) &=& \sqrt{\lambda ({\s}) (1-4\frac{{\ml}^2}{{\s}})},\\
C^{eff}_9({\s}) &=& C_9 +g({\mc},{\s})(3C_1+C_2+3C_3+C_4+3C_5+C_6)-\frac{1}{2}g(1,{\s})(4C_3+4C_4+3C_5+C_6)\nonumber\\
                &-& \frac{1}{2}g(0,{\s})(C_3+3C_4)+\frac{2}{9}(3C_3+C_4+3C_5+C_6),\\
g(0,{\s}) &=& \frac{8}{27}-\frac{8}{9}\ln \frac{m_b}{\mu}-\frac{4}{9}\ln {\s}+\frac{4}{9}i\pi ,\\
g(z,{\s}) &=& \frac{8}{27}-\frac{8}{9}\ln \frac{m_b}{\mu}-\frac{8}{9}\ln z+\frac{4}{9} x \nonumber\\
          &-& \frac{2}{9}(2+x)\sqrt{|1-x|}\left\{
\begin{array}{ll}
(\ln |\frac{\sqrt{1-x}+1}{\sqrt{1-x}-1}|-i\pi),\hspace{0.7cm} for \hspace{0.2cm} x\equiv \frac{4z^2}{{\s}} <1 \\
2\arctan \frac{1}{\sqrt{x-1}},\hspace{1.4cm} for \hspace{0.2cm} x\equiv \frac{4z^2}{{\s}} >1.
\end{array}
\right.
\end{eqnarray}
with
${\s}=\frac{q^2}{m_B^2}$,${\mkks}=\frac{m_{K,K^*}}{m_B}$,${\mbc}=\frac{m_{b,c}}{m_B}$
and ${\ml}=\frac{m_l}{m_B}$. For the Wilson coefficients $C_i$, we
take the results calculated in the naive dimensional
regularization (NRD) scheme \cite{bmmp} and their values are
listed in Table 2.

\begin{center}
\parbox{12cm}{
\small
\baselineskip=1.2pt
 Table 2. Values of the Wilson coefficients with choosing the
 renormalization scale at $\mu=m_b=4.8$GeV. Here,
$C^{eff}_7=C_7-C_5/3-C_6$. }
\end{center}
\begin{center}
\begin{tabular}{|c|c|c|c|c|c|c|c|c|c|}
\hline \hline
\hspace{0.4cm}$C_1$\hspace{0.4cm} & \hspace{0.4cm}$C_2$\hspace{0.4cm} & \hspace{0.4cm}$C_3$\hspace{0.4cm} &
\hspace{0.4cm}$C_4$\hspace{0.4cm} &
\hspace{0.4cm}$C_5$\hspace{0.4cm} & \hspace{0.4cm}$C_6$\hspace{0.4cm} & \hspace{0.4cm}$C_7$\hspace{0.4cm} &
\hspace{0.4cm}$C^{eff}_7$\hspace{0.4cm} &
\hspace{0.4cm}$C_9$\hspace{0.4cm} & \hspace{0.4cm}$C_{10}$\hspace{0.4cm}  \\
\hline
-0.248 & 1.107 & 0.011 & -0.026 & 0.007 & -0.031 & -0.342 & -0.313 & 4.344 & -4.669 \\
\hline \hline
\end{tabular}
\end{center}
\vspace{0.5cm}

It is seen that once the decay rates and the form factors are precisely determined,
one is able to extract the CKM matrix elements. Here we may use the current reasonable values
of CKM matrix elements extracted from other processes and unitarity of the CKM matrix to predict the
branching ratios for the B meson rare decays. When taking the CKM matrix element
$|V_{tb}V_{ts}|=0.0385$, we present our results in Table 3. For comparison, we also list the results
given in \cite{abhh} and the ones from experimental measurements. It is seen that within the uncertainties,
our prediction is consistent with the experimental results.

\begin{center}
\parbox{12cm}{
\small
\baselineskip=1.0pt
 Table 3. Branching ratios ($Br$) for $B$ rare decays in standard model (SM). In deriving the branching ratios we have used
the lifetime of $B$ meson: $\tau_{B}=1.65\mbox{ps}$.  Where the
errors mainly come from the uncertainties of the threshold energy
$s_0$, the Borel parameter $T$ and the parameters in (4.3). }
\end{center}

\begin{center}
\begin{tabular}{|c|c|c|c|}
\hline \hline
 &\hspace{0.7cm} present values \hspace{1.2cm} &\hspace{0.2cm} values in \cite{abhh} \hspace{0.2cm} &
 \hspace{1cm}experiment \hspace{1.5cm} \\
\hline
$B\to K e\bar e$   &                    &                &   \\
                   &(0.84$^{+0.10}_{-0.24}$)$\times 10^{-6}$ & (0.57$^{+0.16}_{-0.10}$)$\times 10^{-6}$
& (0.75$^{+0.25}_{-0.21}\pm 0.09$)$\times 10^{-6}$\cite{kek} \\
$B\to K \mu\bar \mu$ &                    &                                      &   \\
\hline
$B\to K \tau\bar \tau$ &(1.73$^{+0.21}_{-0.47}$)$\times 10^{-7}$ & (1.3$^{+0.3}_{-0.1}$)$\times 10^{-7}$ &  $------$ \\
\hline
                   &           &          &   (4.55$^{+0.72}_{-0.68}\pm 0.34$)$
                   \times 10^{-5 \hspace{0.2cm}\triangle \triangle}$\cite{cleo} \\
$B\to K^* \gamma$ &(5.47$^{+2.14}_{-2.34}$)$\times 10^{-5}$ & (5.37$^{+1.77}_{-1.20}$)$\times 10^{-5 \hspace{0.2cm} \triangle}$
& (4.96$\pm$ 0.67$\pm$0.45)$\times 10^{-5 \hspace{0.2cm} \triangle \triangle}$\cite{belle}   \\
                   &            &           & (5.7 $\pm$ 3.3)$\times 10^{-5}$\cite{group} \\
\hline
$B\to K^* e\bar e$ & (1.86$^{+0.52}_{-0.73}$)$\times 10^{-6}$ & (2.3$^{+0.7}_{-0.4}$)$\times 10^{-6}$
& (2.08$^{+1.23+0.35}_{-1.00-0.37}$)$\times 10^{-6}$\cite{kek}  \\
\hline
$B\to K^* \mu\bar \mu$ &(1.78$^{+0.49}_{-0.70}$)$\times 10^{-6}$ & (1.9$^{+0.5}_{-0.3}$)$\times 10^{-6}$  & $------$   \\
\hline
$B\to K^* \tau\bar \tau$ &(1.68$^{+0.18}_{-0.55}$)$\times 10^{-7}$ & (1.9$^{+0.1}_{-0.2}$)$\times 10^{-7}$ & $------$  \\
\hline \hline
\end{tabular}
\end{center}
\begin{center}
\parbox{12cm}{
\small
\baselineskip=1.0pt
$\triangle$ :The value is obtained through (6.3) by using the form factor $T_1(0)=0.379^{+0.058}_{-0.045}$
presented in Ref.\cite{abhh}.\\
$\triangle \triangle$ : The value is for $B^0 \to {K^0}^*\gamma$.
}
\end{center}

\vspace{1.2cm}

\section{Summary}\label{summary}

 In summary, we have studied the $B$ meson rare decays in the framework of HQEFT. 
The semileptonic type and penguin type form factors have been
derived by using LCSR method in HQEFT. It has been seen that the heavy quark expansion brings a much simplification
to $B$ meson rare decays. Isgur-Wise relations among the semileptonic type and penguin type form factors
have been proved to hold for the whole momentum transfer region at the leading order $1/m_Q$ expansion.
As a consequence, all the form factors can be neatly characterized
by a set of wave functions ($A$, $B$ and $L_i (i=1,2,3,4)$) at the leading order of $1/m_Q$ expansion.
Furthermore from our quantitative discussion, it is obvious that LEET
is a valid method for heavy-to-light transition. LEET relations hold within 80\% accuracy
at large recoil point on the whole, and most of them even hold better than 90\% accuracy.
Moreover, our numerical prediction is consistent with the experimental results.
We then conclude that the branching ratios of $B$ meson rare decays can be reasonably predicted
based on LCSR approach within the framework of HQEFT. Nevertheless, in order to match the expected measurements of B factories in the near
future, a more accurate calculation of the form factors for the B meson rare decays is urgently required.

\acknowledgments

This work was supported in part by the key projects of National Science Foundation
of China (NSFC) and  Chinese Academy of Sciences.

\vspace{2cm}

\newcounter{cankao}
\begin{list}
{[\arabic{cankao}]}{\usecounter{cankao}\itemsep=0cm}
\centerline{\bf REFERENCES}
\vspace{0.3cm}
\small
\bibitem{abhh} A. Ali, P. Ball, L. T. Handoko and G. Hiller, Phys. Rev. D {\bf 61}, 074024 (2000).
\bibitem{akos} T. M. Al\.{i}ev, A. \"{O}zp\.{i}nec\.{i}, M. Savci and H. Koru, Phys. Lett. B {\bf 400}, 194  (hep-ph/9702209).
\bibitem{aos} T. M. Al\.{i}ev, A. \"{O}zp\.{i}nec\.{i} and M. Savci, Phys. Rev. D {\bf 56}, 4260 (1997).
\bibitem{saf} A. S. Safir, EPJ C {\bf 15}, 1 (2001)  (hep-ph/0109232)
\bibitem{ylw} Y. L. Wu, Mod. Phys. Lett. A {\bf 8}, 819 (1993).
\bibitem{wwy} W. Y. Wang, Y. L. Wu and Y. A. Yan, Int. J. Mod. Phys. A {\bf 15}, 1817 (2000).
\bibitem{yww} Y. A. Yan, Y. L. Wu and W. Y. Wang, Int. J. Mod. Phys. A {\bf 15}, 2735 (2000).
\bibitem{ww} W. Y. Wang and Y. L. Wu, Int. J. Mod. Phys. A {\bf 16}, 377 (2001).
\bibitem{bpi} W. Y. Wang, Y. L. Wu, Phys. Lett. B {\bf 515} 57 (2001). (hep-ph/0105154)
\bibitem{brho} W. Y. Wang, Y. L. Wu, Phys. Lett. B {\bf 519} 219 (2001). (hep-ph/0106208)
\bibitem{wwz} W. Y. Wang, Y. L. Wu and M. Zhong, hep-ph/0205157.
\bibitem{iw} N. Isgur and M. B. Wise, Phys. Rev. D {\bf 42}, 2388 (1990)
\bibitem{BD} G. Burdman and J. F. Donoghue, Phys. Lett. B {\bf 270}, 55 (1991).

\bibitem{stech} B. Stech, Phys. Lett. B {\bf 354} 447 (1995).
\bibitem{soares} J. M. Soares, Phys. Rev. D {\bf 54}, 6837 (1996).
\bibitem{ABS} A. Ali, V. M. Braun, H. Simma, Z. Phys. C {\bf 63}, 437 (1994).
\bibitem{cyopr} J. Charles, A. Le Yaouanc, L. Oliver, O. P$\grave{e}$ne and J. C. Raynal, Phys. Rev. D
      {\bf 60}, 014001 (1999). (hep-ph/9812358)
\bibitem{efg} D. Ebert, R. N. Faustov and V. O. Galkin, Phys. Rev. D {\bf 64}, 094022 (2001).
\bibitem{mbtf} M. Beneke and Th. Feldmann, Nucl. Plys. B {\bf 592}, 3 (2001) (hep-ph/0008255)
\bibitem{kms} G. Kramer, G. A. Mannel, T. Schuler, Z. Phys. C {\bf 51}, 649 (1991)
\bibitem{gzmy} G. Burdman, Z. Ligeti, M. Neubert and Y. Nir, Phys. Rev. D {\bf 49}, 2331 (1994).
\bibitem{hly} C. S. Huang, C. Liu and C. T. Yan, Phys. Rev. D {\bf 62}, 054019 (2000).
\bibitem{vai} V. L. Chernyak and I. R. Zhitnitsky, Nucl. Phys. B {\bf 345}, 137 (1990)
\bibitem{ar} A. Khodjamirian and R. R\"uckl, Adv. Ser. Direct. High Energy Phys. {\bf 15}, 345 (1998)
          (WUE-ITP-97-049, MPI-PhT/97-85, hep-ph/9801443).
\bibitem{var} V. M. Belyaev, A. Khodjamirian and R. R\"uckl, Z. Phys. C {\bf 60}, 349 (1993).
\bibitem{pvmisu} P. Ball and V. M. Braun, Phys. Rev. D {\bf 55}, 5561 (1997).
\bibitem{pvesrb} P. Ball and V. M. Braun, Phys. Rev. D {\bf 58}, 094016 (1998). (hep-ph/9805422)
\bibitem{pab} P. Ball, JHEP 09 (1998) 005
\bibitem{cfss} P. Colangelo, F. De Fazio, P. Santorelli, and E. Scrimieri, Phys. Rev. D {\bf 53}, 3672 (1996).
\bibitem{arsco} A. Khodjamirian, R. R\"uckl, S. Weinzierl, C. W. Winhart and O. Yakovlev,
     Phys. Rev. D {\bf 62}, 114002 (2000). (hep-ph/0001297).
\bibitem{vvar} V. M. Belyaev, V. M. Braun, A. Khodjamirian and R. R\"uckl, Phys.Rev.
     D {\bf 51} 6177 (1995).
\bibitem{vi} V. M. Braun and I. B. Filyanov, Z. Phys, C {\bf 44}, 157 (1989).
\bibitem{va} V. L. Chernyak and A. R. Zhitnitsky, Phys. Rep. {\bf 112}, 173 (1984).
\bibitem{bf} V. M. Braun and I. B. Filyanov, Z. Phys. C {\bf 48}, 239 (1990).
\bibitem{pb} P. Ball, JHEP 01 (1999) 010
\bibitem{BH} G. Burdman, G. Hiller, Phys. Rev. D {\bf 63}, 113008 (2001)   (hep-ph/0112063).
\bibitem{MS} D. Melikhov, B. Stech, Phys. Rev. D {\bf 62}, 014006.
\bibitem{UKQCD} UKQCD Collaboration, L. Del Debbio  et al, Phys. Lett. B {\bf 416}, 392 (1998).
\bibitem{ABA} A. Abada, D. Becirevic, Ph. Boucaud, J. P. Leroy, V. Lubicz, G. Martinelli, F. Mescia,
     Nucl. Phys. Proc. Suppl. {\bf 83}, 268 (2000)    (hep-lat/9910021).
\bibitem{TAI} Chuan-Hung Chen, C. Q. Geng, Nucl. Phys. B {\bf 636}, 338   (hep-ph/0203003).
\bibitem{bm} A. J. Buras, M. M\"unz, Phys. Rev. D {\bf 52}, 186 (1995)
\bibitem{bmmp} A. J. Buras, M. Misiak, M. M\"unz and S. Pokorski, Nucl. Phys. B {\bf 424}, 374 (1994)
\bibitem{kek} Belle Collaboration, K. Abe et al., Phys. Rev. Lett. {\bf 88(2)},021801(6) (2002)
\bibitem{cleo} CLEO Collaboration, T. E. Coan et al. Phys. Rev. Lett. {\bf 84}, 5283 (2000)
\bibitem{belle} G. Taylor,[BELLE Collaboration], talk presented at the 36th Rencontres de Moriond
                Electroweak Interactions and Unified Theories, Les Arcs, France, March 2001.
\bibitem{group} Particle Data Group, D. E. Groom et al, EPJ C {\bf 15} 1 (2000)
\end{list}
\newpage
\centerline{\large{FIGURES}}
\vspace{1.0cm}
\newcommand{\PICL}[2]
{
\begin{center}
\begin{picture}(460,150)(0,0)
\put(0,40){
\epsfxsize=7.0cm
\epsfysize=4.2cm
\epsffile{#1} }
\put(100,14){\makebox(0,0){#2}}
\end{picture}
\end{center}
}

\newcommand{\PICR}[2]
{
\begin{center}
\begin{picture}(300,0)(0,0)
\put(160,66){
\epsfxsize=7.0cm
\epsfysize=4.2cm
\epsffile{#1} }
\put(275,40){\makebox(0,0){#2}}
\end{picture}
\end{center}
}
\small
\mbox{}
\vspace{1cm}

\PICL{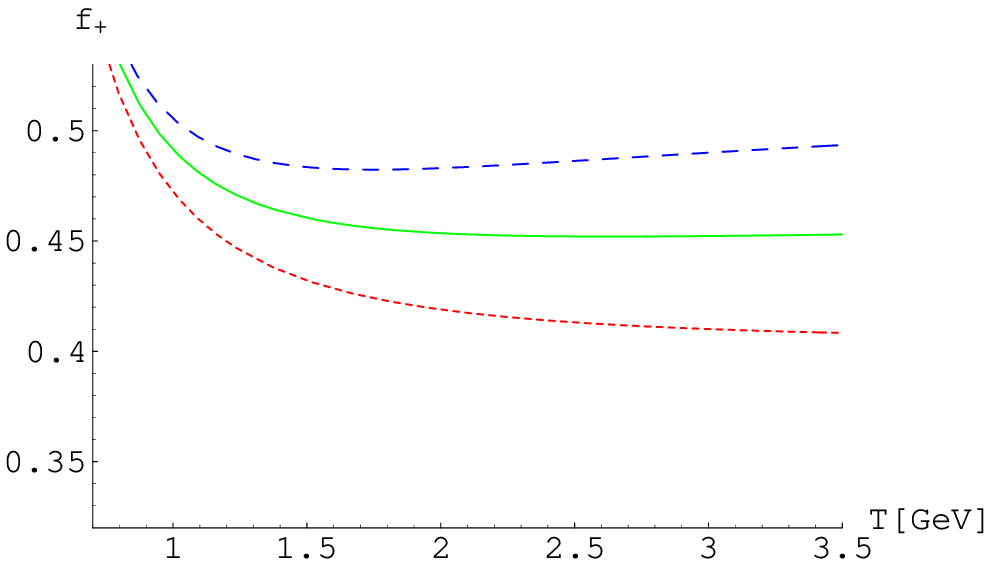}{\centerline{
\parbox{7cm}{
\small
\baselineskip=1.0pt
Fig.1 The form factors $f_+$ as functions of
the Borel parameter $T$ for different values of the threshold $s_0$. The dotted, solid
and dashed curves correspond to $s_0=2.1,2.4$ and 2.7GeV respectively.
The momentum transfer is $q^2=0$.}}}

\PICR{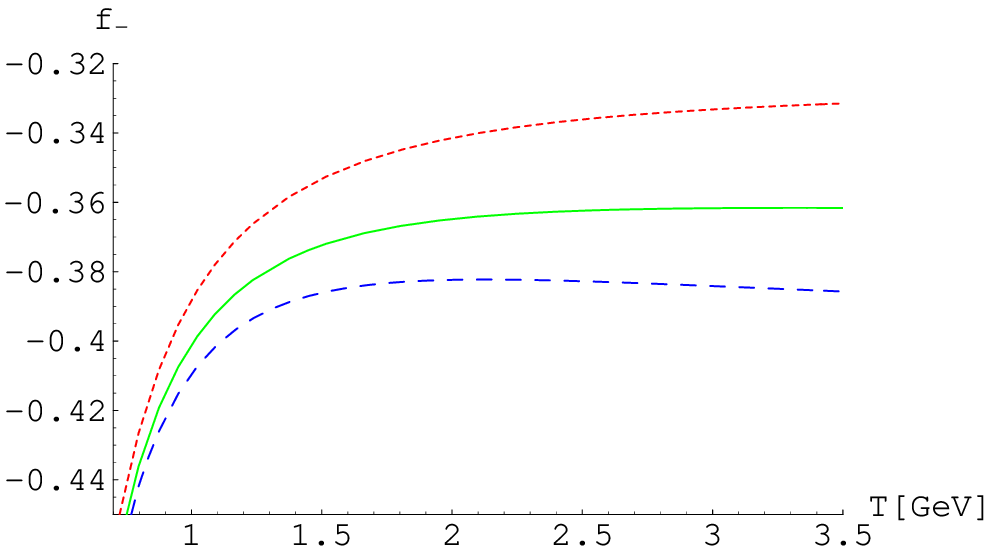}{Fig.2 Same as Fig.1 but for $f_-$.}

\PICL{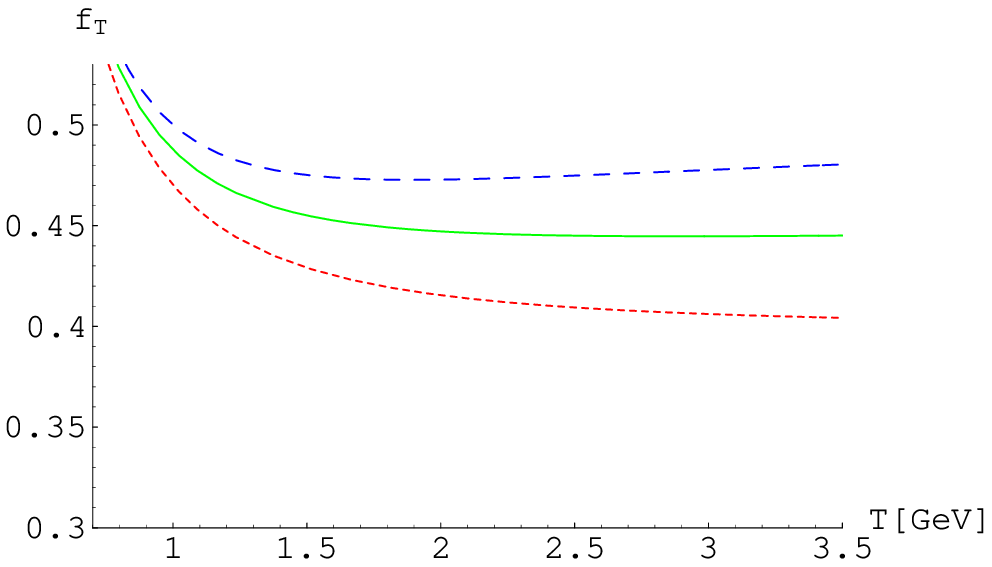}{Fig.3 Same as Fig.1 but for $f_T$.}

\PICR{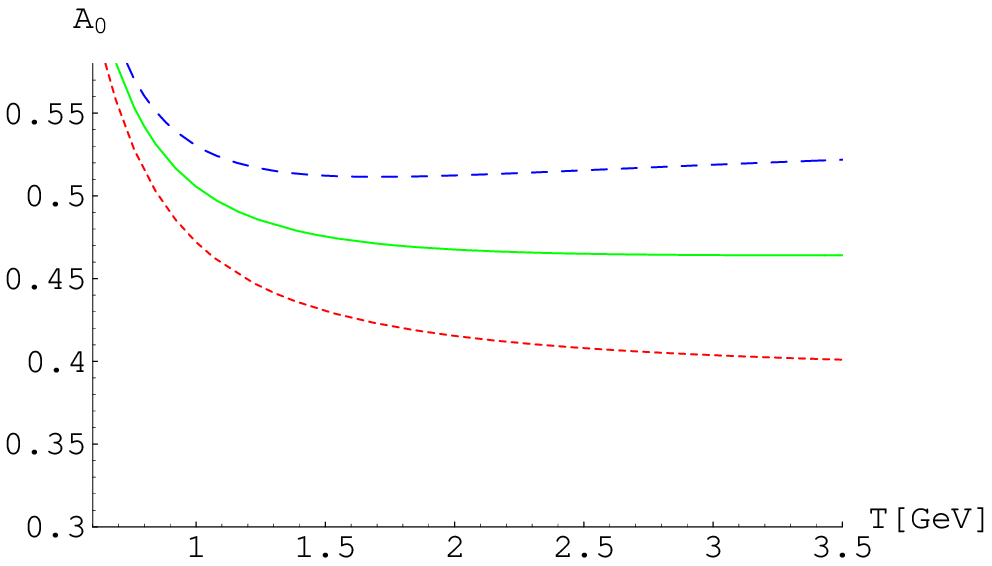}{\centerline{
\parbox{7cm}{
\small
\baselineskip=1.0pt
Fig.4 The form factors $A_0$ as functions of
the Borel parameter $T$ for different values of the threshold $s_0$. The dotted, solid
and dashed curves correspond to $s_0=1.8,2.1$ and 2.4GeV respectively.
The momentum transfer is $q^2=0$.}}}

\PICL{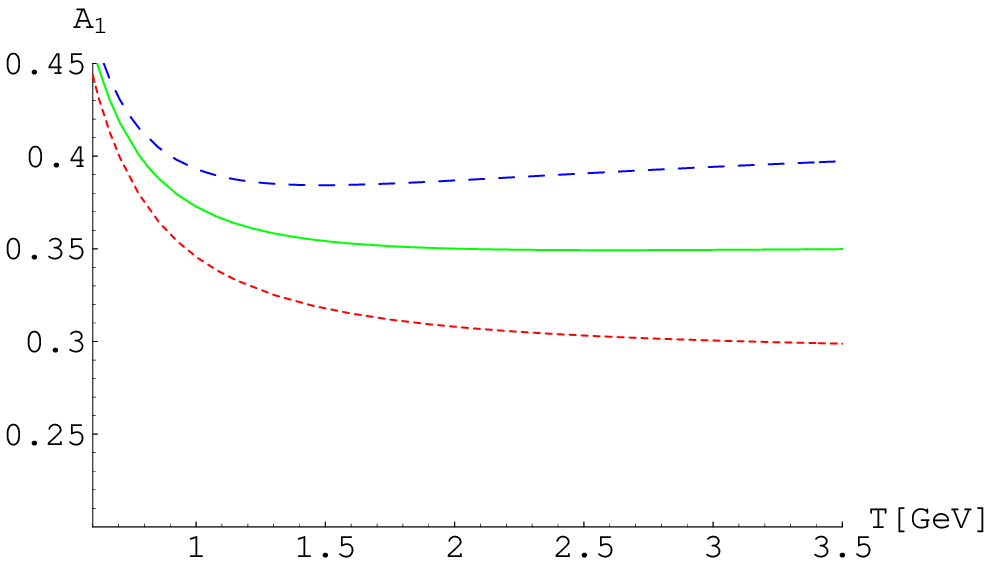}{Fig.5 Same as Fig.4 but for $A_1$.}

\PICR{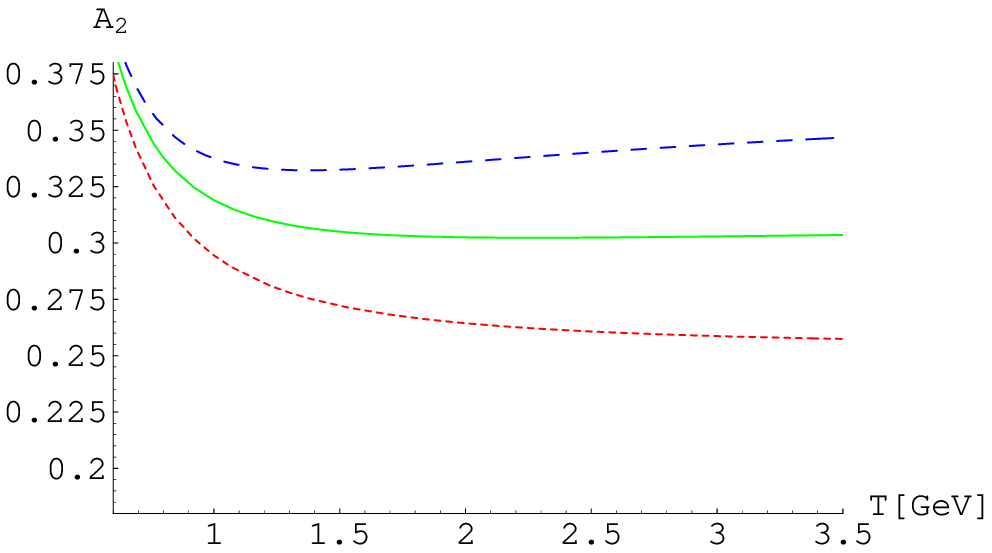}{Fig.6 Same as Fig.4 but for $A_2$.}

\PICL{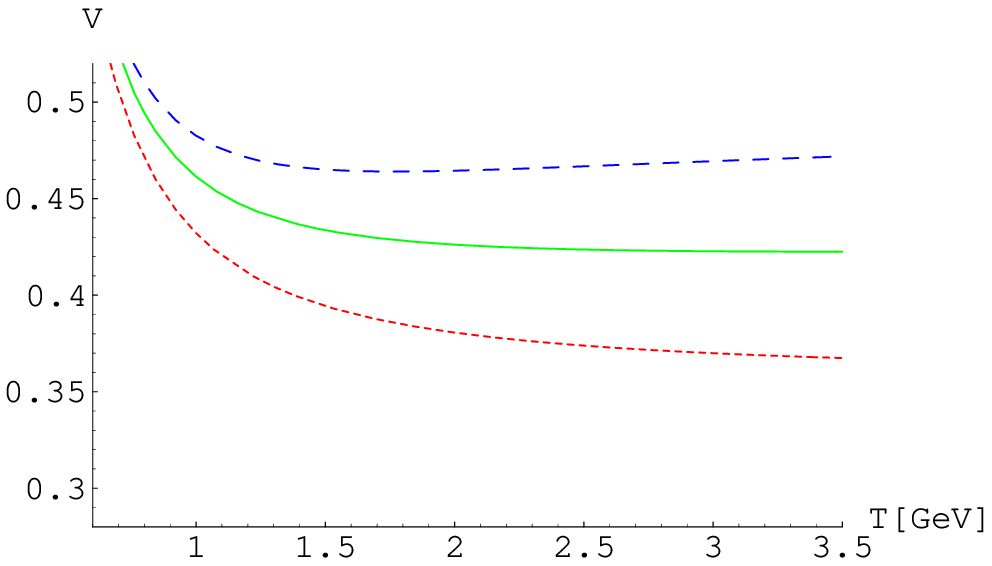}{Fig.7 Same as Fig.4 but for $V$.}

\PICR{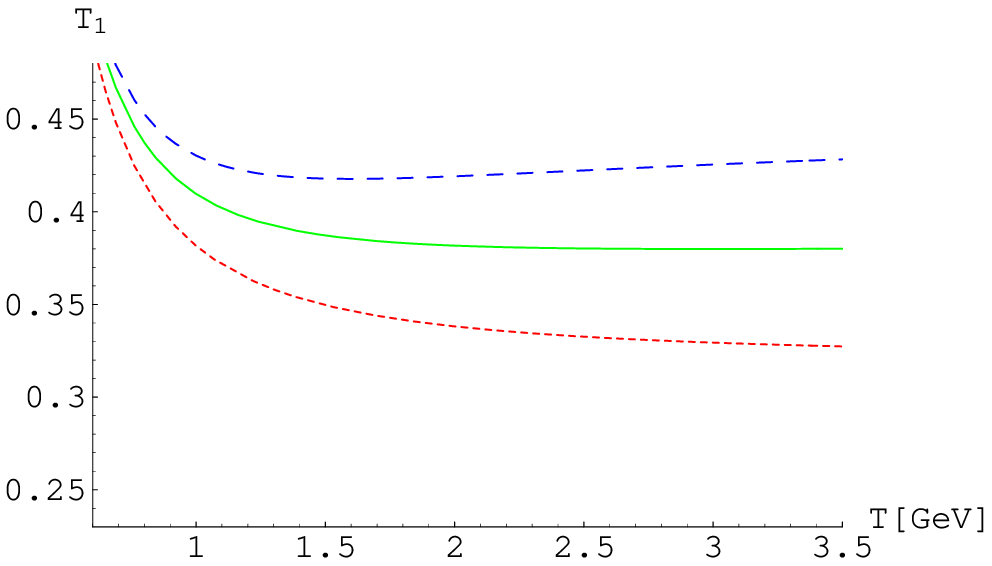}{Fig.8 Same as Fig.4 but for $T_1$.}

\PICL{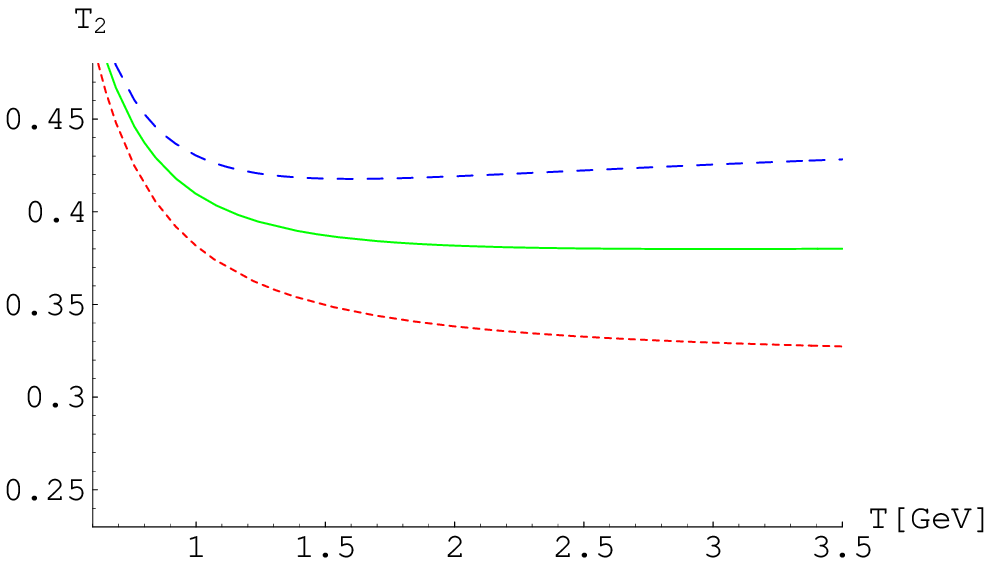}{Fig.9 Same as Fig.4 but for $T_2$.}

\PICR{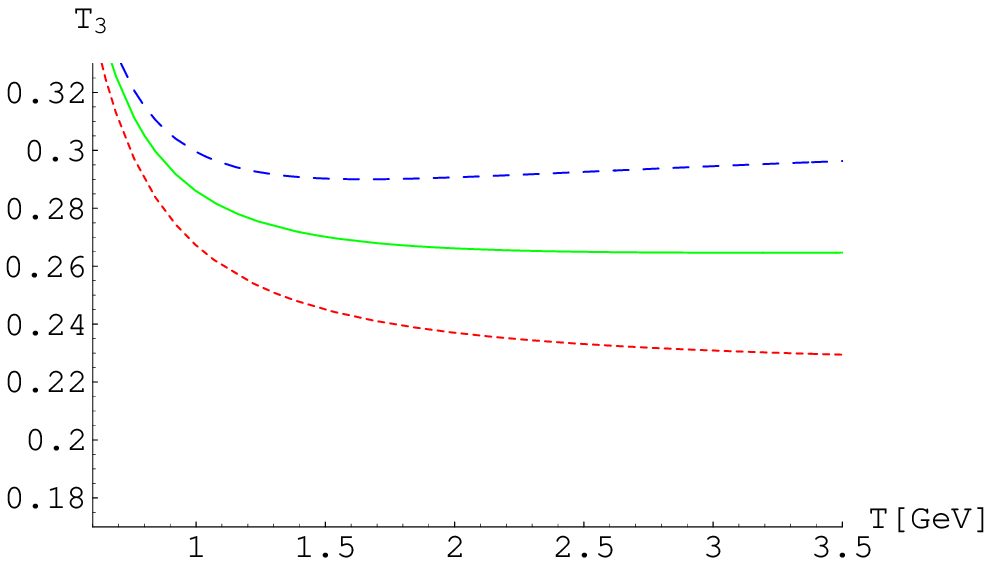}{Fig.10 Same as Fig.4 but for $T_3$.}

\PICL{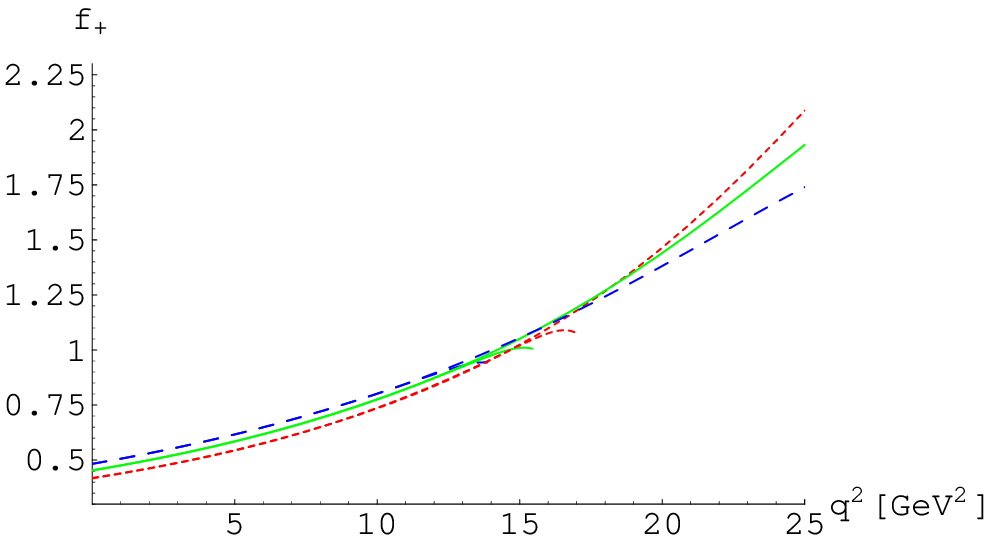}{\centerline{
\parbox{7cm}{
\small
\baselineskip=1.0pt
Fig.11 The form factors $f_+$ as functions of
 $q^2$ for different values of the threshold $s_0$. The dotted, solid
and dashed curves correspond to $s_0=2.1, 2.4$ and 2.7GeV respectively.
}}}

\PICR{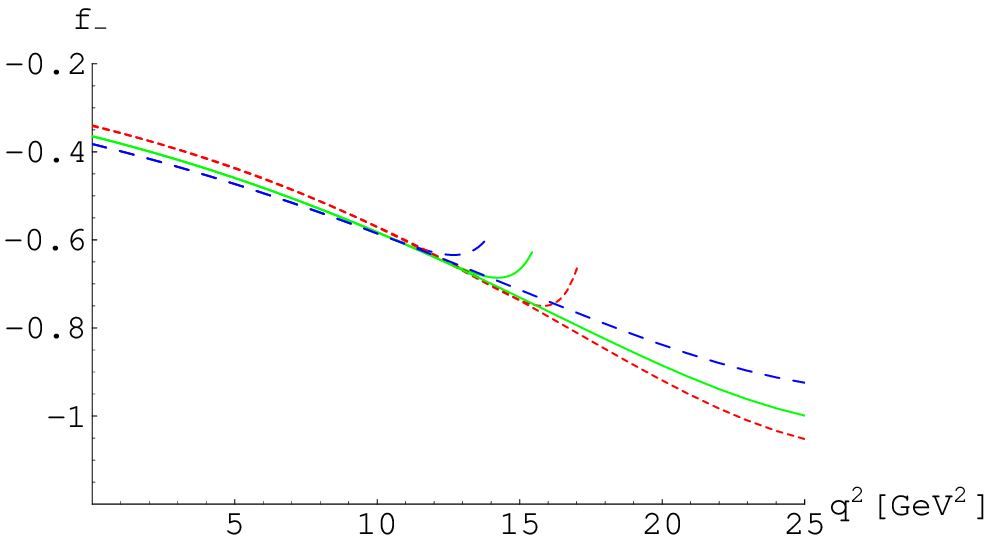}{Fig.12 Same as Fig.11 but for $f_-$.}

\PICL{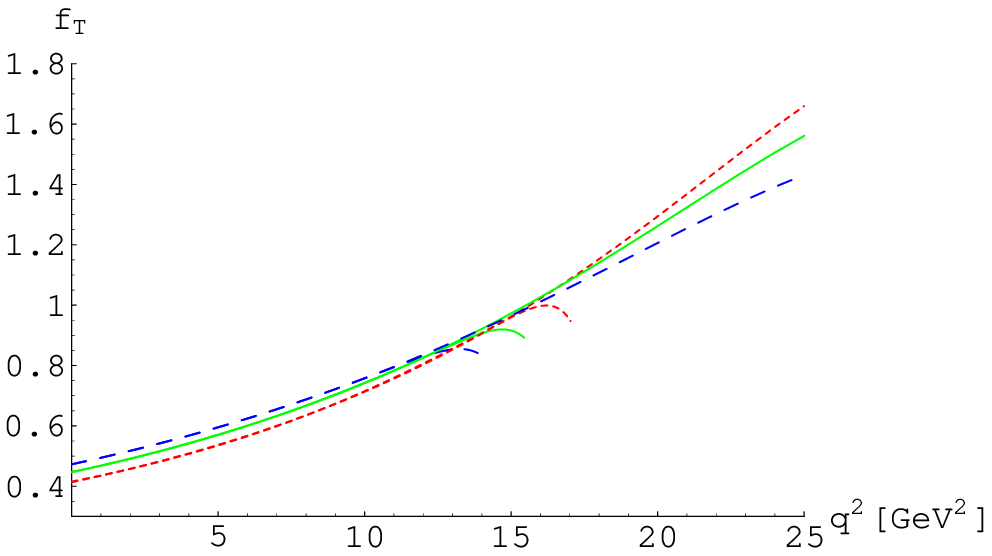}{Fig.13 Same as Fig.11 but for $f_T$.}
\PICR{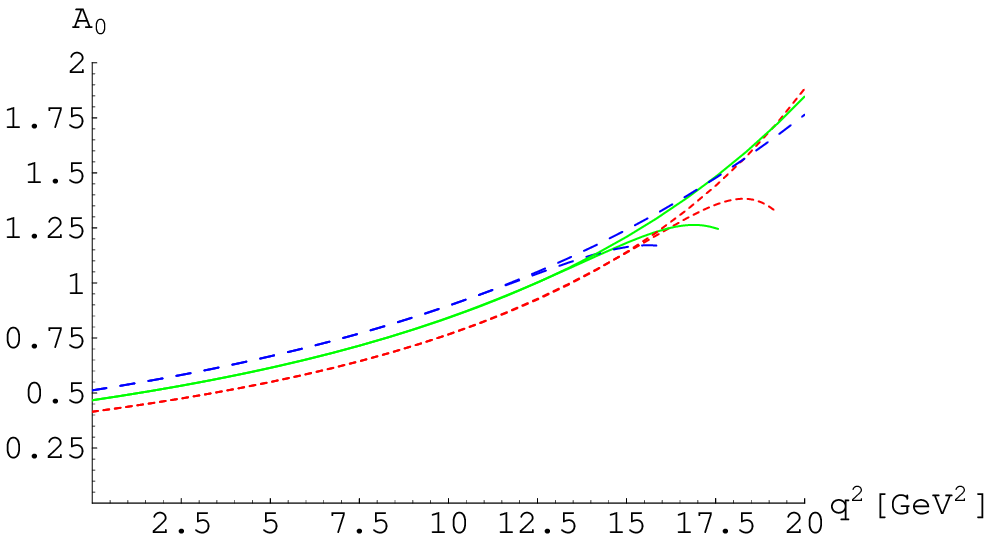}{\centerline{
\parbox{7cm}{
\small
\baselineskip=1.0pt
Fig.14 The form factors $A_0$ as functions of
$q^2$ for different values of the threshold $s_0$. The dotted, solid
and dashed curves correspond to $s_0=1.8, 2.1$ and 2.4GeV respectively.
}}}

\PICL{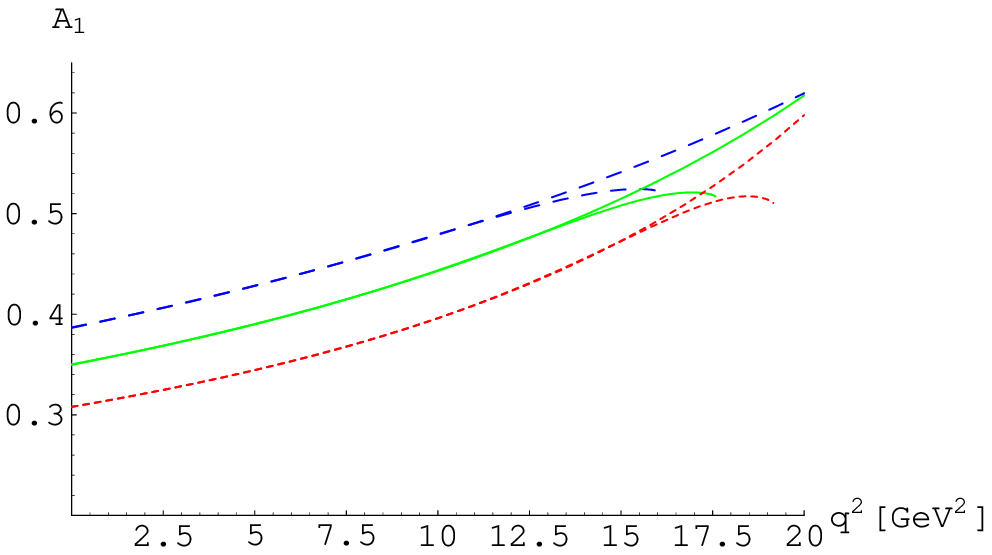}{Fig.15 Same as Fig.14 but for $A_1$.}

\PICR{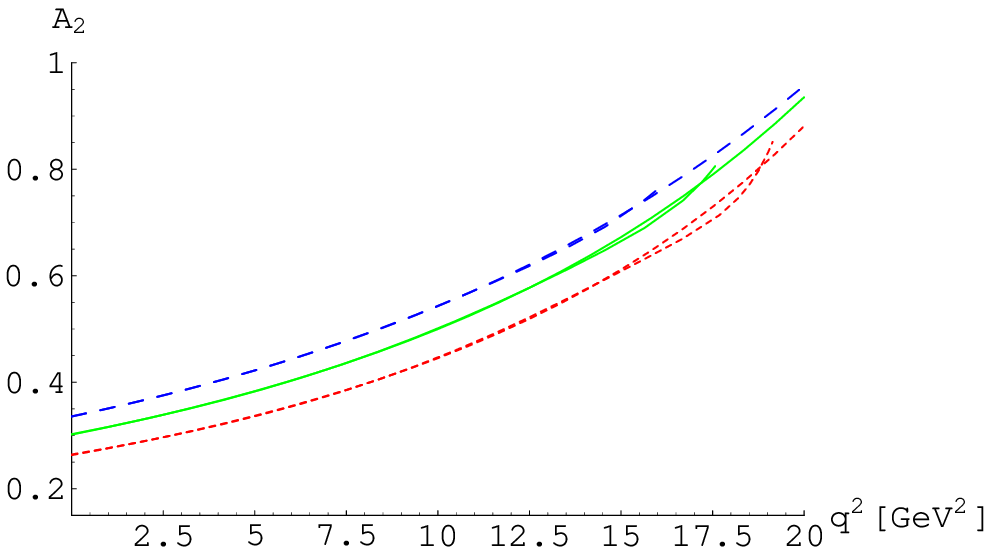}{Fig.16 Same as Fig.14 but for $A_2$.}

\PICL{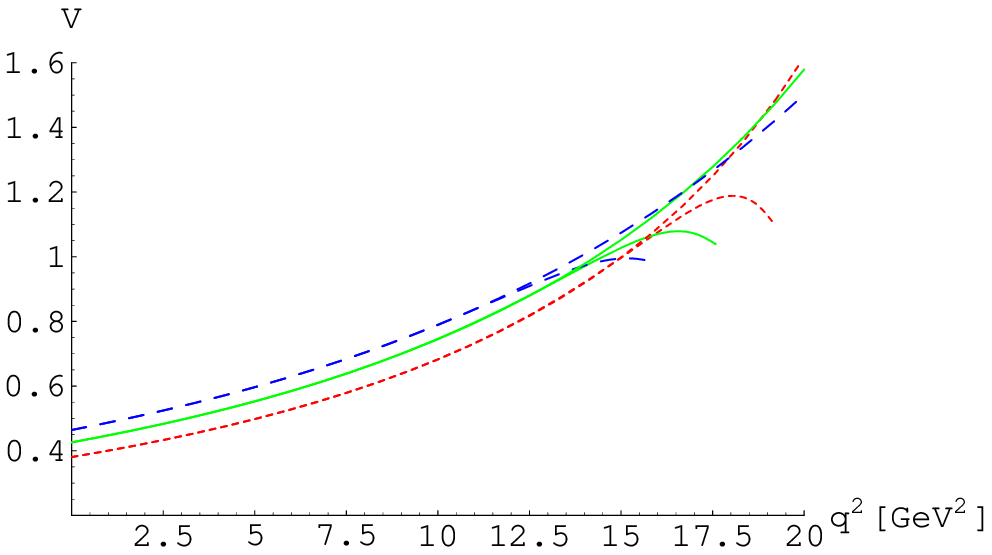}{Fig.17 Same as Fig.14 but for $V$.}

\PICR{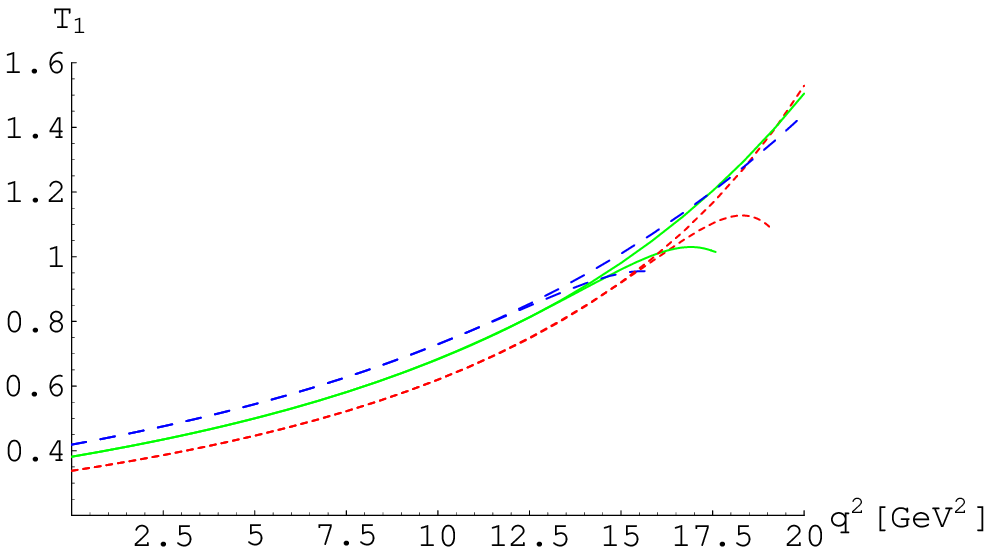}{Fig.18 Same as Fig.14 but for $T_1$.}

\PICL{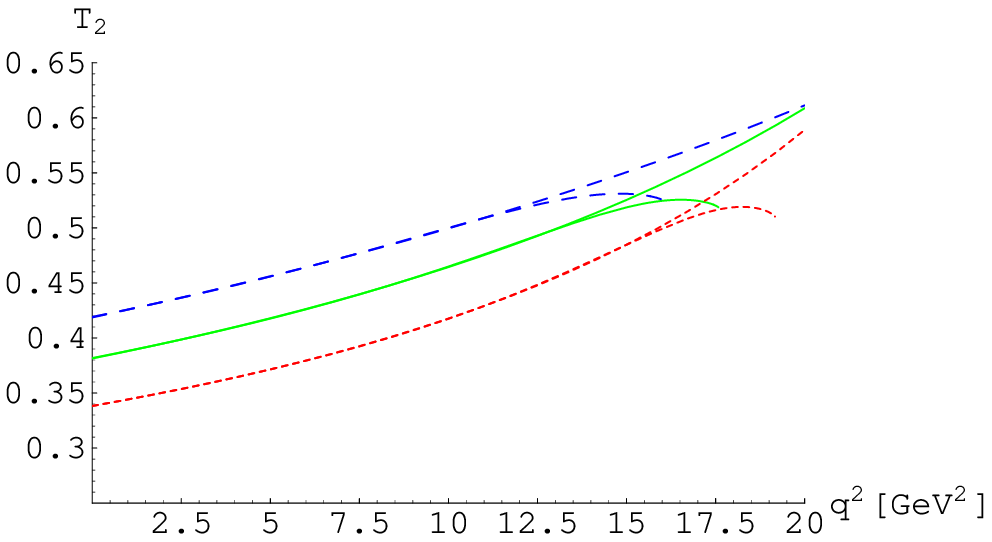}{Fig.19 Same as Fig.14 but for $T_2$.}

\PICR{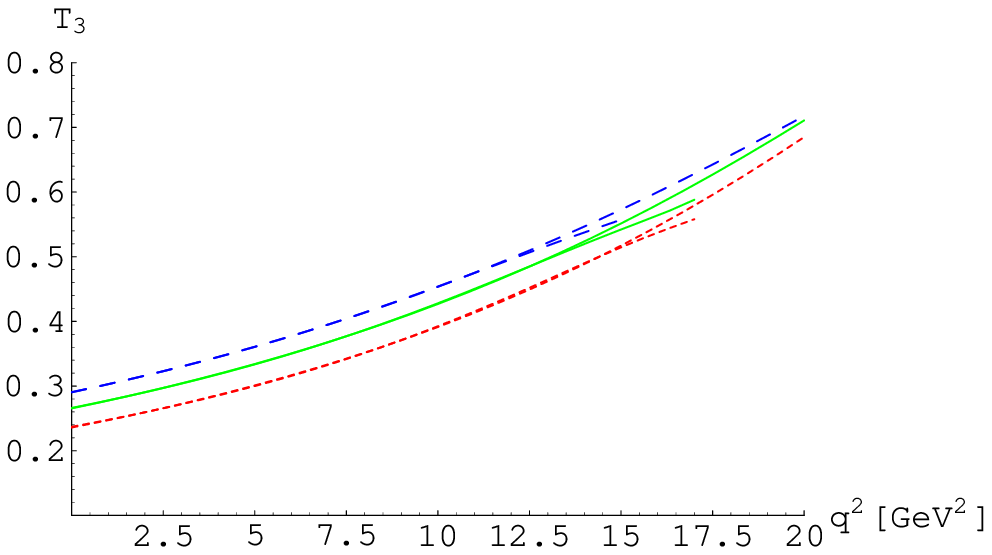}{Fig.20 Same as Fig.14 but for $T_3$.}

\PICL{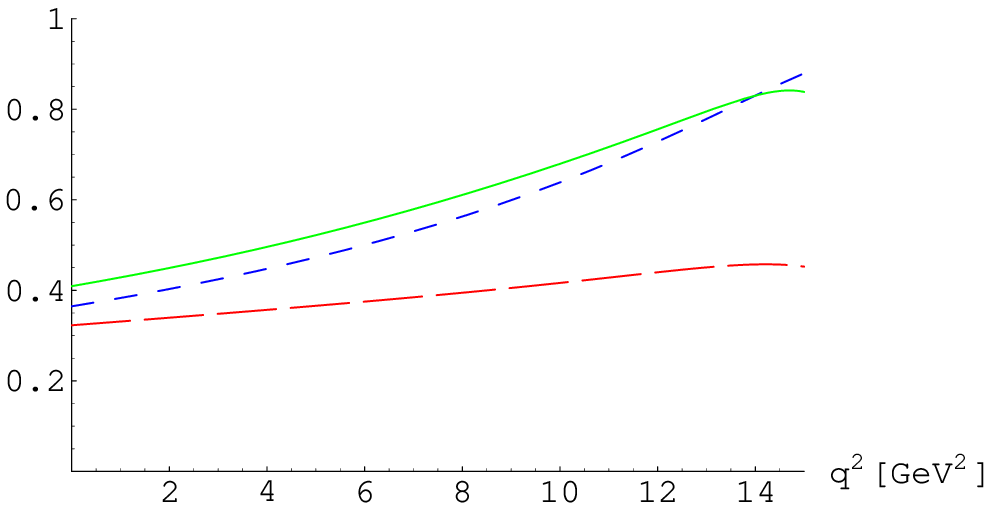}{\centerline{
\parbox{6cm}{
\small
\baselineskip=1.0pt
Fig.21 $\zeta_{\parallel}(q^2)$, $\zeta_{\perp}(q^2)$ and $\zeta(q^2)$ at center value of $s_0$ with $T=2.0$GeV.
The dotted, dashed and solid curves correspond to
$\zeta_{\parallel}(q^2)$, $\zeta_{\perp}(q^2)$ and $\zeta(q^2)$ respectively.}}}

\PICR{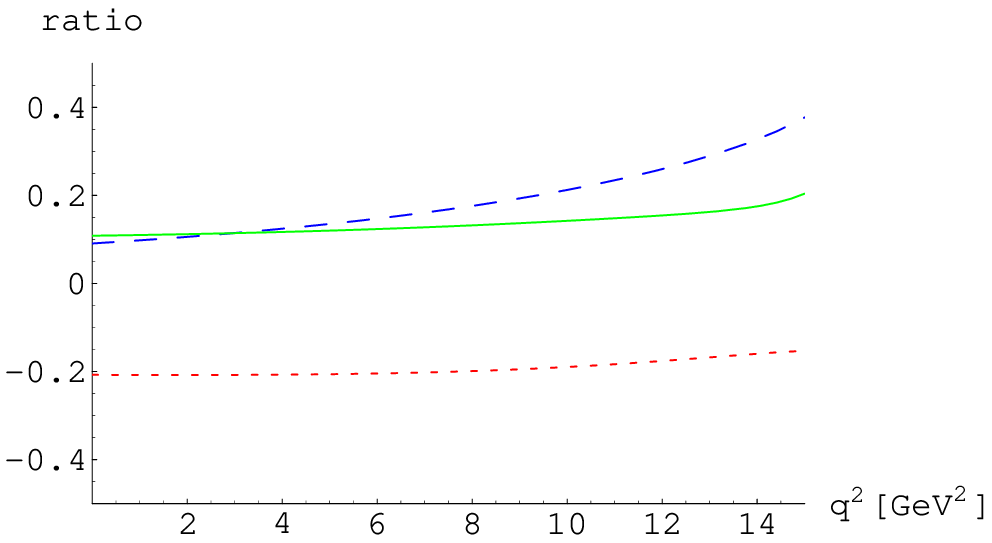}{\centerline{
\parbox{7.3cm}{
\small
\baselineskip=1.0pt
Fig.22 $\zeta_2(q^2)/\zeta_{\parallel}(q^2)$, $\zeta_1(q^2)/\zeta_{\perp}(q^2)$ and $\zeta_A(q^2)/\zeta(q^2)$
at center value of $s_0$ with $T=2.0$GeV.
The dotted, dashed and solid curves correspond to
$\zeta_2(q^2)/\zeta_{\parallel}(q^2)$, $\zeta_1(q^2)/\zeta_{\perp}(q^2)$ and $\zeta_A(q^2)/\zeta(q^2)$ respectively.}} }

\PICL{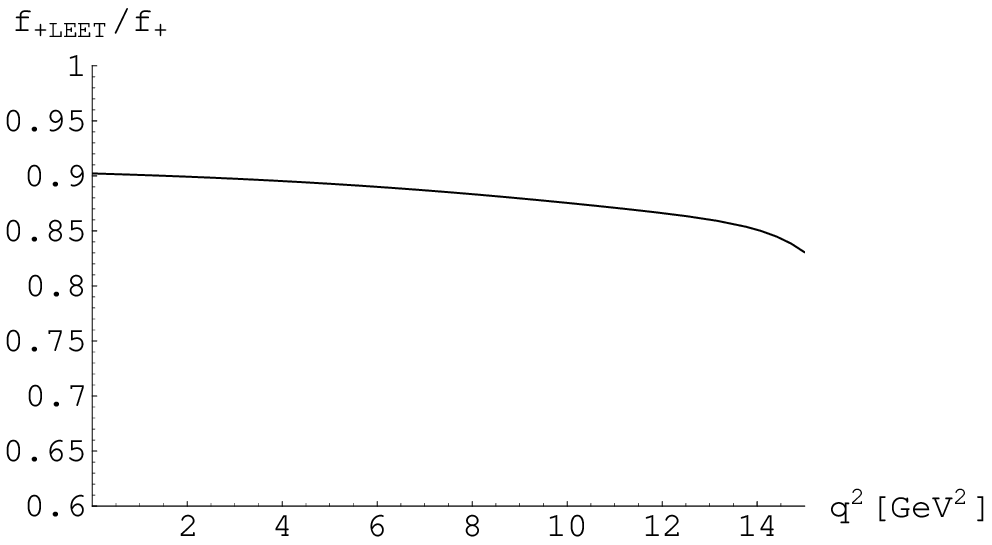}{\centerline{
\parbox{5cm}{
\small
\baselineskip=1.0pt
Fig.23 $f_{+ LEET}(q^2)/f_+(q^2)$ at $s_0=2.4$GeV with $T=2.0$GeV.}} }

\PICR{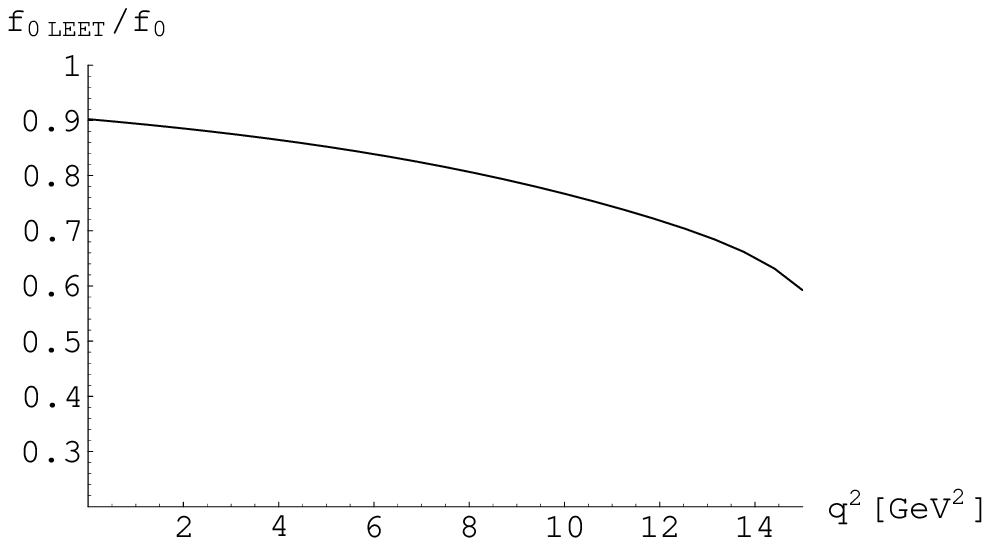}{\centerline{
\parbox{5cm}{
\small
\baselineskip=1.0pt
Fig.24 Same as Fig.23 but for $f_{- LEET}(q^2)/f_-(q^2)$.}} }

\PICL{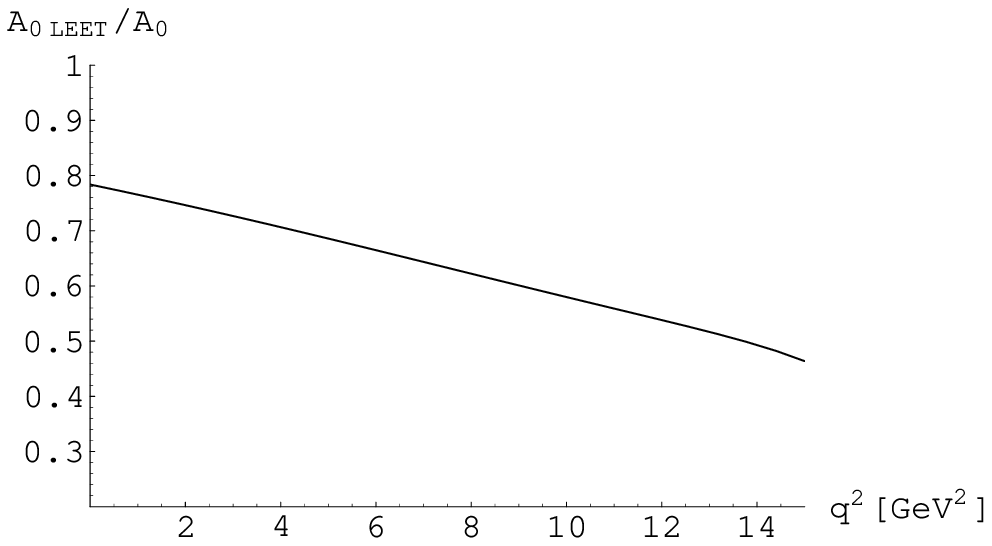}{\centerline{
\parbox{5cm}{
\small
\baselineskip=1.0pt
Fig.25 $A_{0 LEET}(q^2)/A_0(q^2)$ at $s_0=2.1$GeV with $T=2.0$GeV.}} }

\PICR{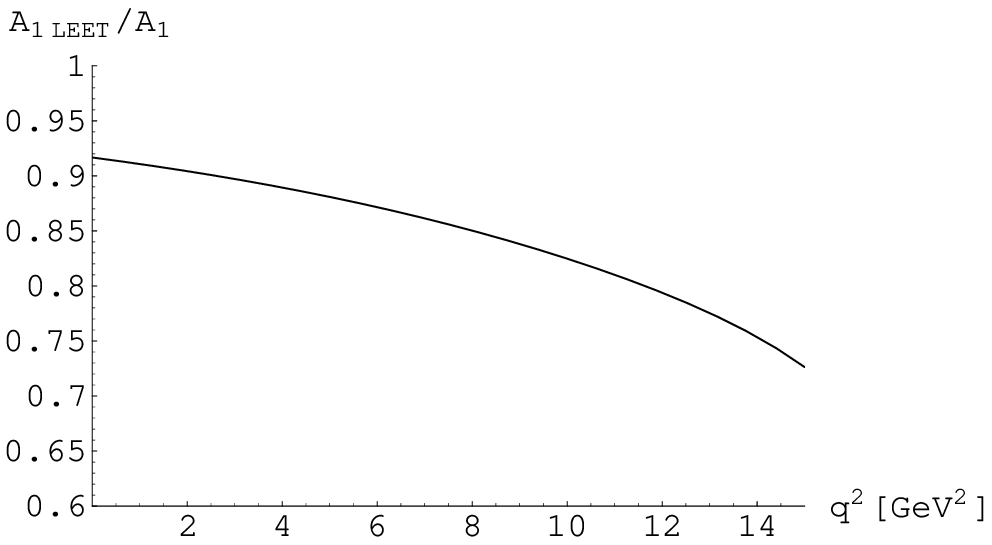}{\centerline{
\parbox{5cm}{
\small
\baselineskip=1.0pt
Fig.26 Same as Fig.25 but for $A_{1 LEET}(q^2)/A_1(q^2)$.}} }

\PICL{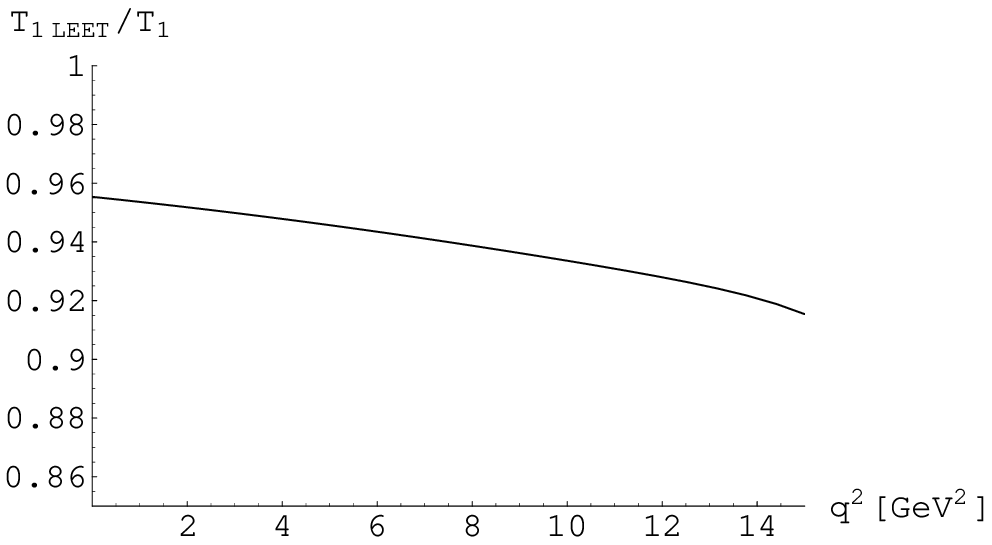}{\centerline{
\parbox{5cm}{
\small
\baselineskip=1.0pt
Fig.27 Same as Fig.25 but for $T_{1 LEET}(q^2)/T_1(q^2)$.}} }

\PICR{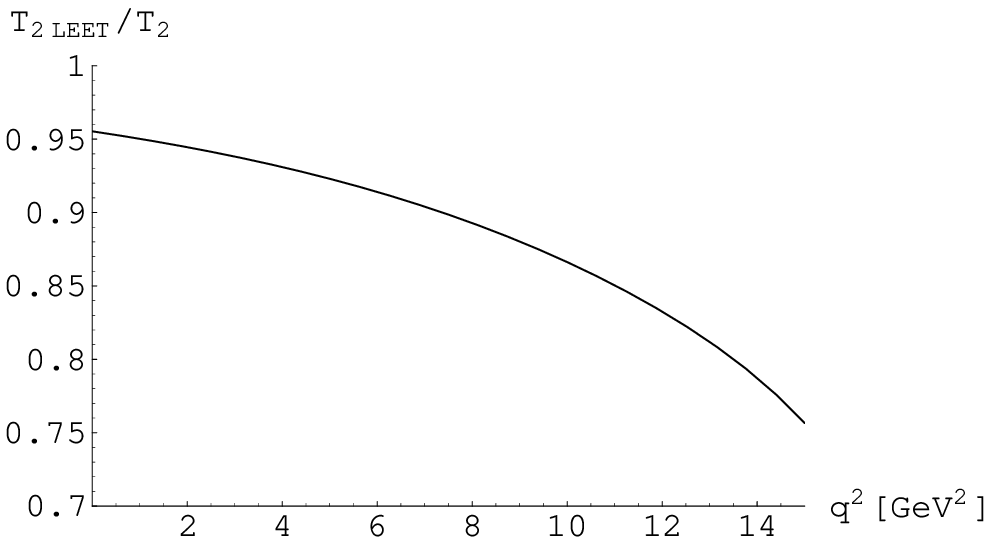}{\centerline{
\parbox{5cm}{
\small
\baselineskip=1.0pt
Fig.28 Same as Fig.25 but for $T_{2 LEET}(q^2)/T_2(q^2)$.}}}

\PICL{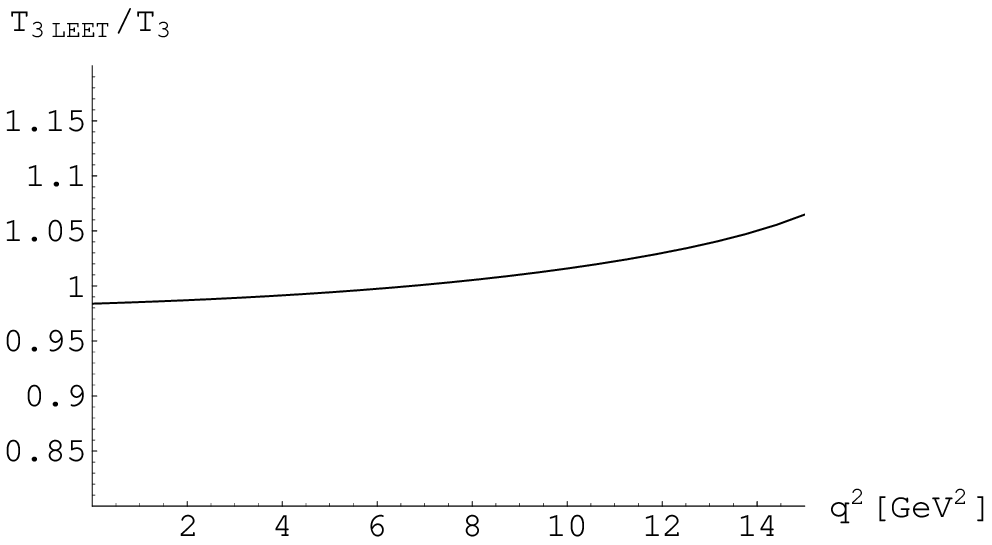}{\centerline{
\parbox{5cm}{
\small
\baselineskip=1.0pt
Fig.29 Same as Fig.25 but for $T_{3 LEET}(q^2)/T_3(q^2)$.}} }

\PICR{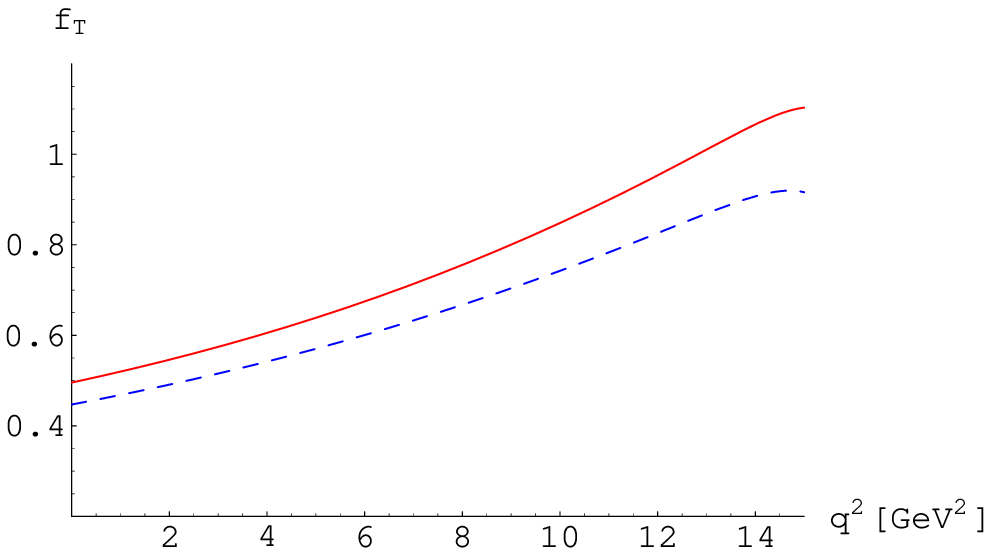}{\centerline{
\parbox{6cm}{
\small
\baselineskip=1.0pt
Fig.30 LEET relation (5.17) for $B\to K$ at $s_0=2.4$GeV and $T=2.0$GeV.
The dashed and solid curves correspond to LHS and RHS of the equation respectively.}}}

\PICL{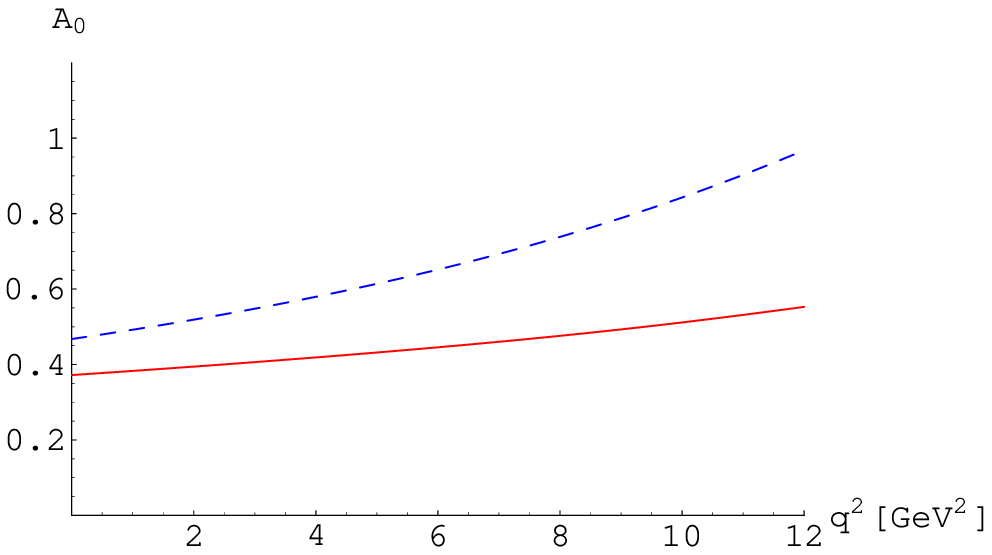}{\centerline{
\parbox{6cm}{
\small
\baselineskip=1.0pt
Fig.31 LEET relation (5.18) for $B\to K^*$ at $s_0=2.1$GeV and $T=2.0$GeV.
The dashed and solid curves correspond to LHS and RHS of the equation respectively.}} }

\PICR{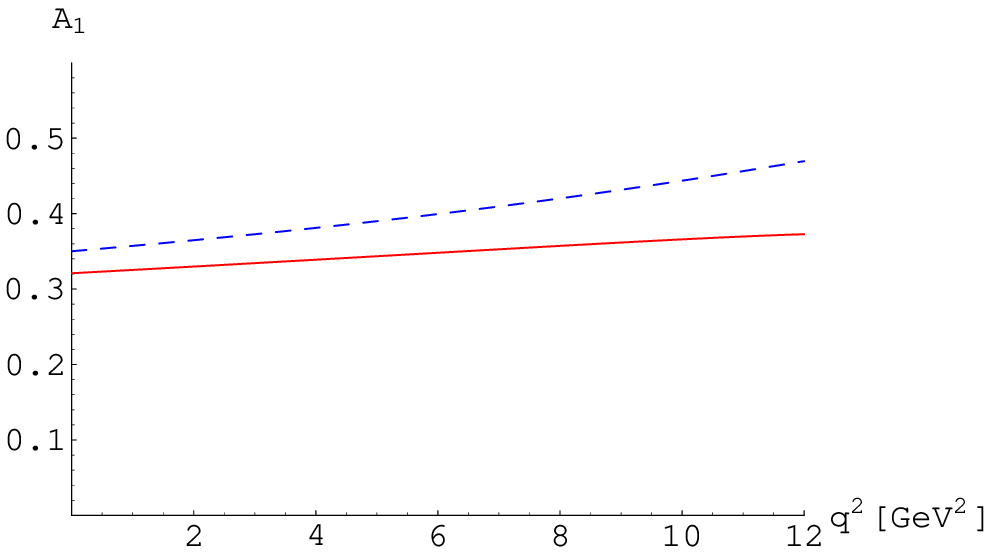}{\centerline{
\parbox{6cm}{
\small
\baselineskip=1.0pt
Fig.32 LEET relation (5.19) for $B\to K^*$ at $s_0=2.1$GeV and $T=2.0$GeV.
The dashed and solid curves correspond to LHS and RHS of the equation respectively.}} }
\end{document}